\documentclass[CRPHYS,Unicode,manuscript]{cedram}

\usepackage{graphicx}  
\usepackage{dcolumn}   
\usepackage{bm}        
\usepackage{amssymb}   
\usepackage{amsmath,epsfig}   
\usepackage{verbatim}  
\usepackage{cases}
\usepackage{mathrsfs} 
\usepackage{color}

\definecolor{darkgreen}{rgb}{0.0, 0.5, 0.0}

\usepackage{xcolor}
\hypersetup{
    colorlinks,
    linkcolor={red!50!black},
    citecolor={blue!50!black},
    urlcolor={blue!80!black}
}

\newcommand{\bml}{\begin{multline}}
\newcommand{\bea}{\begin{eqnarray}}
\newcommand{\eea}{\end{eqnarray}}
\newcommand{\be}{\begin{equation}}
\newcommand{\ee}{\end{equation}}
\newcommand{\bi}{\begin{itemize}}
\newcommand{\ei}{\end{itemize}}
\newcommand{\ds}{\displaystyle}
\newcommand{\rr}{\mathbf{r}}
\newcommand{\kk}{{\mathbf{k}}}
\newcommand{\KK}{{\mathbf{K}}}

\newcommand{\uu}{\mathbf{u}}
\newcommand{\RR}{\mathbf{R}}
\newcommand{\LL}{\mathbf{L}}
\newcommand{\CC}{\mathbf{C}}
\newcommand{\cc}{\mathbf{c}}
\newcommand{\JJ}{\mathbf{J}}

\newcommand{\XX}{\mathbf{X}}

\newcommand{\vn}{\mathbf{0}}
\newcommand{\ra}{\rangle}
\newcommand{\la}{\langle}

\newcommand{\ktyp}{k_{\rm typ}}

\newcommand{\gr}{\bm \nabla}

\newcommand{\UP}{\uparrow}
\newcommand{\down}{\downarrow}

\newcommand{\Ar}{\mathcal{A}}

\newcommand{\Vr}{\mathcal{V}}

\newcommand{\Cr}{\mathcal{C}}
\newcommand{\Fr}{\mathcal{F}}
\newcommand{\Rr}{\mathcal{R}}

\newcommand{\Wr}{\mathcal{W}}

\newcommand{\Er}{\mathcal{E}}
\newcommand{\Sr}{\mathcal{S}}
\newcommand{\Nr}{\mathcal{N}}
\newcommand{\Or}{\mathcal{O}}
\newcommand{\Mr}{\mathcal{M}}

\newcommand{\Oo}{{\bm \Omega}} 
\newcommand{\rrho}{\bm \rho} 
\newcommand{\md}{\mathsf{m}} 
\newcommand{\dK}{d^3\!K}
\newcommand{\dr}{d^3\!r}
\newcommand{\dc}{d^3\!c}
\newcommand{\dC}{d^3\!C}

\newcommand{\bl}{} 

\newcommand{\dd}{d_2}
\newcommand{\ddd}{{\bl d_3}}

\newcommand{\gre}{\bl} 
\newcommand{\brown}{\bl} 

\newcommand{\bcbp}{}

\begin{document}


\title{Three-body contact for fermions. \\I. General relations}

\author{\firstname{F\'elix} \lastname{Werner} \CDRorcid{0000-0002-5631-9024}}
\address{Laboratoire Kastler Brossel, Ecole Normale Sup\'erieure - Universit\'e PSL, CNRS, Coll\`ege de France, Sorbonne Universit\'e, 75005 Paris, France}
\email{werner@lkb.ens.fr}

\author{\firstname{Xavier} \lastname{Leyronas} \CDRorcid{0000-0002-9499-6800}}
\address{Laboratoire de Physique de l'Ecole Normale Sup\'erieure, ENS - Universit\'e PSL, Sorbonne Universit\'e, Universit\'e Paris Cité, CNRS, 75005 Paris, France}
\email{xavier.leyronas@phys.ens.fr}

\date{\today}

\begin{abstract}
  We consider the resonant Fermi gas, that is,
  two-component
  fermions in three dimensions interacting by a short-range potential of large scattering length. We introduce a quantity, the three-body contact, that determines several observables.
Within the zero-range model, the number of nearby fermion triplets, the large-momentum tail of the center-of-mass momentum distribution of  nearby fermion pairs, as well as the large-momentum tail of the two-particle momentum distribution,
are expressed in terms of the three-body contact.
For a small finite interaction range, the formation rate of deeply bound dimers by three-body recombination, as well as the three-body contribution to the finite-range correction to the energy, are expressed in terms of the three-body contact and of a three-body parameter. This three-body parameter, which vanishes in the zero-range limit, is defined through the asymptotic behavior of the zero-energy scattering state at distances intermediate between the range and the two-body scattering length. In~general, the three-body contact has different contributions labeled by spin and angular momentum indices,
and the three-body parameter can depend on those indices.
We also include the generalization to unequal masses for $\uparrow$ and $\downarrow$ particles. With respect to the relation between three-body loss rate and number of nearby triplets stated in [Petrov, Salomon and Shlyapnikov, PRL {\bf 93}, 090404 (2004)], the present work adds a derivation, expresses the proportionality factor in terms of the three-body parameter, and includes the general case where there are several contributions to the three-body contact and several three-body parameters.
\end{abstract}

\begin{altabstract}
Nous considérons le gaz de Fermi résonnant, à savoir des fermions avec deux états internes à trois dimensions avec des interactions à courte portée de grande longueur de diffusion. Nous introduisons une quantité, le contact à trois corps, qui détermine plusieurs observables. Pour le modèle de portée nulle, le nombre de triplets de fermions proches, la queue de la distribution selon l'impulsion du centre de masse des paires de fermions proches, ainsi que la queue de la distribution en impulsion à deux particules, sont exprimées en termes du contact à trois corps. Pour une portée non nulle, le taux de formation de dimères fortement liés par recombinaison à trois corps, ainsi que la contribution à trois corps à la correction de portée finie à l'énergie, sont exprimées en termes du contact à trois corps et d'un paramètre à trois corps. Ce paramètre à trois corps, qui tend vers zéro dans la limite de portée nulle, est défini {\it via} le comportement asymptotique de l'état de diffusion d'énergie nulle à des distances intermédiaires entre la portée et la longueur de diffusion à deux corps. En général, le contact à trois corps a différentes contributions repérées par des indices de spin et de moment cinétique, et le paramètre à trois corps peut dépendre de ces indices. Nous incluons aussi la généralisation à des masses différentes pour les particules $\uparrow$ et $\downarrow$. Par rapport à la relation donnée dans [Petrov, Salomon et Shlyapnikov, PRL {\bf 93}, 090404 (2004)] entre taux de pertes à trois corps et nombre de triplets de fermions proches, le présent travail ajoute une dérivation, exprime le facteur de proportionnalité en termes du paramètre à trois corps, et inclus le cas général où il y a plusieurs contributions au contact à trois corps et plusieurs paramètres à trois corps.
\end{altabstract}
\maketitle


\section{Introduction}

Over the last twenty years, the two-component Fermi gas with zero-range interactions in three dimensions has become one of the most extensively studied quantum many-body problems.
{\bl One considers particles with two internal states
(denoted $\UP$ and $\down$)
and an interaction of  vanishing range
characterized by its $s$-wave scattering length $a_2$.}
When $1/a_2$ changes from $-\infty$ to $+\infty$, the interaction changes from weakly to strongly attractive, leading to the BCS to BEC crossover.
The strongly correlated regime is reached in the central region of the crossover, around the unitary limit $1/a_2=0$.
While the model was historically introduced as a theoretical abstraction~\cite{Leggett_Chap_1980_book,Haussmann_Z_Phys,Haussmann_PRB},
it accurately describes ultracold gases of fermionic atoms in two hyperfine states near a Feshbach resonance,
which are the subject of numerous experimental studies, see {\it e.g.}~\cite{RevueTrentoFermions,ZwergerBook2,ChapZwergerZwerger,ThomasStronglyInteractingGaz,GrimmCrossover,JinPairCondensate,KetterlePairCondensate,SalomonCrossover,GrimmModes,ShinPhaseDiag,SylEOS,NirEOS,KuEOS,LENSRepulsivePolaron,ValeGoldstone,EsslingerWiedemannFranz,Boiling_MIT,ZwierleinSound,SagiPolaronMol,MoritzBragg3D,NavonRecomb,HuletClosedChannel,AustraliensC,AustraliensT,ValeC_precise,JinUnivRel,Jin_C_homogeneous,SagiContactRF,ValeC_T_hom,ZwierleinC,LaurentC}.

From a theoretical viewpoint, this resonant Fermi gas is a
difficult problem.
As for most strongly correlated many-body problems  in dimension $>1$,
 numerical methods are generally the only option to make precise predictions.
 Furthermore the zero-range nature of the interactions typically
 constitutes
 an additional difficulty for numerical computations.
But zero-range interactions
also give rise to specific exact relations, called Tan relations, involving a ubiquitous quantity, the
two-body contact $C_2$~\cite{TanEnergetics,TanLargeMomentum,ChapLeggettBref,LeChapitreIn2,ChapBraatenBref,RanderiaRF_arxiv,ZwergerRFLong,RanderiaRF,BaymRF,BraatenC,TanViriel,FelixViriel,ZwergerRF,BraatenLong,WernerTarruellCastin,ZhangLeggettUniv,CombescotC,WernerCastinRelationsArxiv,HofmannAnomaly,TanTrapFunctional,Moelmer,WernerCastinRelationsFermions}.
In particular, $C_2$ determines the probability to find two particles close to each other.
For the resonant Fermi gas,
$C_2$ was measured and computed in numerous studies,
see {\it e.g.}~\cite{WernerTarruellCastin,HuletClosedChannel,AustraliensC,AustraliensT,ValeC_precise,JinUnivRel,NirEOS,Jin_C_homogeneous,SagiContactRF,ValeC_T_hom,ZwierleinC,LaurentC}
and~\cite{Baym_C,Hu_C,SunVirial3,StrinatiC,ZwergerViscosity,Carlson_C_relations,Drut_C,ValeC_precise,WetterichC,Goulko_UFG_2016,RossiContact,AlhassidC} respectively.

Similarly, universal relations involving a two-body contact $C_2$ hold
for two-component fermions in 2D~\cite{FelixViriel,CombescotC,WernerCastinRelationsArxiv,Moelmer,BraatenRFshift2D,HofmannAnomaly,WernerCastinRelationsFermions,MolmerAnyD}
and 1D~\cite{FelixViriel,ZwergerRelations1D},
and for single-component bosons in 1D~\cite{FelixViriel,Olshanii_nk}
and 2D~\cite{FelixViriel,WernerCastinRelationsBosons}.
For a 2D Bose gas, $C_2$ was recently measured interferometrically~\cite{beugnonC}.
{\bl
Several two-body contacts appear in the general relations
for single-component fermions with $p$-wave~\cite{UeadeRelationsPwave,ZhangThywissenRelations_pwave,ZhouRelationsLwave,PengLiuHuRelations3Dpwave,ThywissenRelationspwave_exp} or higher partial-wave~\cite{ZhangRelations_dwave} short-range interactions in~3D,
and dipolar plus short-range interactions in 2D~\cite{HofmannDipolarRelations}.}

For bosons in 3D {\bl with resonant interactions}, 
in addition to the two-body contact $C_2$, measured in~\cite{jin_contact_BEC,HazibabicC3}, a three-body contact $C_3$ 
{\bl appears in} several exact relations~\cite{FelixViriel,WernerCastinRelationsArxiv,BraatenBosons,CastinWerner_nk_trimer,WernerCastinRelationsBosons}.
{\bl This appearance of $C_3$ is linked to the Efimov effect. In particular, a three-body parameter has to be included in the definition of the zero-range model, and $C_3$ is proportional to the derivative of the energy w.r.t. the three-body parameter.} 
$C_3$~was measured interferometrically in~\cite{HazibabicC3}
after an interaction quench to unitarity.
{\bl A three-body contact also appears for single-component fermions
  with higher-partial-wave resonant interactions, both in 2D~\cite{ZhangRelations2Dpwave_superefi} (where a super-Efimov effect occurs) and in 1D~\cite{NishidaRelations1D,NishidaRelations1D_2}.}
{\bl Two-body and three-body contacts were also found to be useful to describe short-distances or large-momenta properties in clusters of helium atoms~\cite{BazakHeliumContact} 
and in nuclei~\cite{BarneaNuclContactsPRL2015,BarneaNuclContactsPRC2015,HenPLB2018,Nucl_C_exp_Nature2020,HenContactsNatPhys2021,CLAScollab_PLB2021,GandolfiC3_nucl},   
although the corresponding relations are only approximate because the interaction range is not much smaller than the interparticle distance.}

{\bl Here we show that a three-body contact $C_3$ plays an important role 
  for two-component fermions with resonant interactions in 3D, although there is no three-body Efimov effect so that the zero-range model is parameterized by the scattering length $a_2$ without any three-body parameter~\cite{Efimov,Petrov3fermions}.}
{\bl We work within the zero-range model in Section~\ref{sec:ZR}, and we consider models with a small finite interaction range in Section~\ref{sec:finite_b}.}
{\bl Within the zero-range model,} the number of triplets of particles separated by a small distance (Section~\ref{sec:N3}), the tail of the center-of-mass momentum distribution of pairs separated by a small distance (Sec.~\ref{sec:tail_NP}), {\gre and the tail of the two-particle momentum distribution (Sec.~\ref{sec:N_k1k2})} are expressed in terms of $C_3$,
and $C_3$ is also related to the third order density correlation function (Sec.~\ref{sec:g3})
and to the behavior of the many-body wavefunction when three particles approach each other (Sec.~\ref{sec:psi}).
{\bl When the interaction range $b$ is non-zero but still small compared to the other typical lengthscales,
we consider two additional observables,}
the formation rate of deeply bound dimers by three-body recombination {\bl $\Gamma_3$} (Sec.~\ref{sec:loss}),
{\bl and the three-body contribution to the energy correction induced by the finite interaction range $\delta\!E_3$ (Sec.~\ref{sec:dE}).
We express $\Gamma_3$ and $\delta\!E_3$ in terms of $C_3$, and of a three-body parameter~$a_3$ (which is small in the zero-range regime).
We define~$a_3$ through the asymptotic behavior of the three-body zero-energy scattering state at distances $\gg b$ and $\ll |a_2|$.}


We consider $N_\UP$ fermions of spin $\UP$ and
$N_\down$ fermions of spin $\down$,
either confined by a smooth external trapping potential,
or in a box with periodic boundary conditions.
{\bl We consider equal masses for $\UP$ and $\down$ particles,
and discuss the unequal-mass case in Appendix~\ref{app:imbal}.}
We consider a stationary state throughout the article,
and discuss statistical mixtures and non-stationary states in Appendix~\ref{app:non_stat}.
In parallel with presenting the relations involving the three-body contact,
we will recall for comparison the known Tan relations involving the two-body contact.

{\bl \section{Relations for the zero-range model} \label{sec:ZR}

In this Section we work within the zero-range model, where interactions are characterized by a single parameter, the two-body scattering length $a_2$.
The zero-range model is defined in Eqs.~(\ref{eq:schro},\ref{eq:BP}).
The zero-range limit of finite-range models is expected to be universally described by the zero-range model.\footnote{{\bl The absence of $(N_\UP+N_\down)$-body Efimov effect (in the equal-mass case) was shown for $(N_\UP,N_\down)=(2,1)$~\cite{Efimov,Petrov3fermions}, $(2,2)$~\cite{Petrov4body2004,EndoCastin2+2,Michelangeli2plus2,Seiringer2plus2}, $(3,1)$~\cite{CMP},
$(4,1)$~\cite{PetrovBazak5body}, $(5,1)$~\cite{Bazak6body}, and $(N,1)$ for any $N$~\cite{SeiringerNplus1}. The zero-range model was proven to be self-adjoint in the $(2,2)$~\cite{Seiringer2plus2} and $(N,1)$~\cite{SeiringerNplus1} cases. In the latter case, some rigorous results about the Tan relations were also obtained~\cite{SeiringerNplus1}.} {\gre The convergence of finite-range models towards the zero-range model in the zero-range limit was confirmed by various theoretical studies, see {\it e.g.} \cite{BlumeRevue,LudoYvanBoite} for single-channel models and \cite{BraatenLong,WernerTarruellCastin,ZhangLeggettUniv,MoraCastinPricoupenkoCRAS,castin_tignone} for two-channel models, and by numerous theory-experiment comparisons for the many-body problem, {\it e.g.} \cite{Giorgini,GrimmModes,ShinPhaseDiag,SylEOS,NirEOS,GezerlisXi,CarlsonAFQMC,ValeC_precise,VanHouckeEOS,KuEOS,RossiEOS,RossiContact,ValeC_T_hom,ZwierleinC,AlhassidC}.}}
}


\subsection{Number of  nearby fermion triplets} \label{sec:N3}

If one measures the positions of all particles, the average number of pairs of particles 
whose separation is smaller than some $\epsilon$ is given by
\be
N_2(\epsilon) \underset{\epsilon\to0}{\sim} \,C_2\ \frac{\epsilon}{4\pi}
\label{eq:N2}
\ee
where $C_2$ is the two-body contact~\cite{TanEnergetics,TanLargeMomentum}.
Similarly, let us consider the number of triplets of fermions separated by small distances.
For three particles 1,2,3,
let us define the hyperradius
\be
R=\sqrt{\frac{2}{3}
\left(r_{12}^{\phantom{aa}2}+r_{13}^{\phantom{aa}2}+r_{23}^{\phantom{aa}2}\right)
\label{eq:def_R}
}
\ee
where $r_{ij}$ is the distance between particles $i$ and $j$.
If one measures the positions of all particles, the average number of triplets of particles with hyperradius $R<\epsilon$ is given by
\be
\boxed{N_3(\epsilon) 
\underset{\epsilon\to0}{\sim} \ C_3\ \epsilon^{2s+2}}
\label{eq:N3}
\ee
where the prefactor $C_3$ is what we call the 
three-body contact,
while the exponent
\\$s=1.772724267\ldots$ is the lowest positive solution different from 1 of
\be
(1-s^2)\,\sin\left(\frac{s\pi}{2}\right)
-\frac{4}{\sqrt{3}}\,s\,\cos\left(\frac{s\pi}{6}\right)
+ 4\,\sin\left(\frac{s\pi}{6}\right)
=0.
\label{eq:transc_s}
\ee

The scaling 
$N_3(\epsilon) \propto \epsilon^{2s+2}$
{\bl was already obtained in~\cite{Petrov4body2004,TanScalingREVTEX}
(see Section~\ref{sec:psi} for a rederivation).}
The anomalous exponent $2s+2$ comes from the analytical solution of the unitary three-body problem~\cite{Efimov},
and is directly linked to a hidden dynamical symmetry
and a separability of the three-body problem in hyperspherical coordinates~\cite{LeChapitreIn2,CRAS,WernerSym},
or in a field theory point of view, to
non-relativistic conformal invariance, with $s+5/2$ the scaling dimension of a three-fermion operator~\cite{ChapNishidaSon,son_cft,Mehen_cft}.

In general there are two different contributions to the three-body contact,
coming from $\UP\UP\down$ and~$\UP\down\down$ spin configurations:
Denoting by
$N_{2,1}$ (resp.~$N_{1,2}$)
the contributions to $N_3(\epsilon)$
from triplets of particles of~spins~$\UP\UP\down$ (resp.~$\UP\down\down$),
we have
\be
\boxed{N_{2,1}(\epsilon) 
\underset{\epsilon\to0}{\sim} \,C_{2,1}\ \epsilon^{2s+2}}
\label{eq:N21}
\ee
\be
\boxed{N_{1,2}(\epsilon) 
\underset{\epsilon\to0}{\sim} \,C_{1,2}\ \epsilon^{2s+2}}
\label{eq:N12}
\ee
{\bl where $C_{2,1}$ (resp.~$C_{1,2}$) is what we call the $\UP\UP\down$ (resp.~$\UP\down\down$) three-body contact.
Clearly,}
\be
\boxed{C_3 = C_{2,1}+C_{1,2}.}
\label{eq:C_3_21_12}
\ee

\vskip 1cm
\noindent \underline{\it Remarks:}
\bi
\item
  Due to the antibunching effect associated to the Pauli exclusion between fermions with identical spins,
the contribution to Eq.~(\ref{eq:N2}) coming from pairs of particles with identical spins ($\UP\UP$~or~$\down\down$) is negligible in the $\epsilon\to0$ limit
(it scales as $\epsilon^5$), and $N_2(\epsilon)$ is dominated by the contribution from pairs of particles with opposite spins ($\UP\down$).
\\Similarly, the contribution to Eq.~(\ref{eq:N3}) coming from triplets of particles with identical spins ($\UP\UP\UP$~or~$\down\down\down$) is negligible in the $\epsilon\to0$ limit
(it scales as $\epsilon^{10}$),
and $N_3(\epsilon)$ is dominated by the contributions from triplets of particles with non-identical spins, $\UP\UP\down$ or $\UP\down\down$,
in agreement with Eq.~(\ref{eq:C_3_21_12}).

\item
For comparison, in the
non-interacting case, 
the number of nearby pairs and triplets scales as
\bea
N_2^{(0)}(\epsilon) &\propto& \epsilon^3
\label{eq:N2_0}
\\
N_3^{(0)}(\epsilon) &\propto& \epsilon^8
\label{eq:N3_0}
\eea
(more generally, these scalings also hold with a finite-range interaction that does not diverge too strongly at small distance, so that the wavefunction is bounded).
\\Equation~(\ref{eq:N2_0}) [resp. Equation~(\ref{eq:N3_0})] is dominated
by the contribution from pairs (resp. triplets) of particles with non-identical spins.
\\Equation~(\ref{eq:N3_0}) includes the antibunching effect due to the Pauli exclusion between the two identical-spin fermions:
The wavefunction vanishes linearly with the distance between these fermions,
hence an $\epsilon^2$ suppression factor compared to the
completely uncorrelated case of non-interacting distinguishable particles
\be
N^{(0)}_{3,\ \rm distinguishable}(\epsilon) \propto \epsilon^6.
\label{eq:N3_dist}
\ee

\item Compared to the non-interacting case Eqs.~(\ref{eq:N2_0},\ref{eq:N3_0}), the exponents in Eqs.~(\ref{eq:N2},\ref{eq:N3})
are reduced, {\it i.e.} the probability to find particles near to each other is enhanced.
This bunching effect is due to the attractive effect of the resonant zero-range interaction,
which causes the wavefunction
to diverge:
When the distance $r$ between two opposite-spin particles vanishes,
 $\psi \propto 1/r$,
which yields Eq.~(\ref{eq:N2}),
while in the limit of vanishing hyperradius $R$ between three particles,
$\psi \propto R^{s-2}$,
which yields Eq.~(\ref{eq:N3}).
Note that since $s<2$, the wavefunction indeed diverges for $R\to0$,
and the exponent in Eq.~(\ref{eq:N3}) is
smaller than in the uncorrelated case Eq.~(\ref{eq:N3_dist}),
which means that the bunching effect due to the zero-range interactions overcompensates the antibunching effect due to Pauli exclusion.\footnote{On the other hand, 
the Pauli exclusion effect is not overcompensated by too much:
Due to the repulsive effective three-body potential $s^2/R^2$
(see App.~\ref{app:unit_hyp}),
we still have
$N_3(\epsilon) / \epsilon^2 \propto \epsilon^{2s} \underset{\epsilon\to0}{\to} 0$, which implies that the three-body loss rate divided by the thermalization rate vanishes in the zero-density limit, a crucial ingredient for the zero-range model to be an accurate description of ultracold-atom experiments (see Sec.~\ref{sec:loss}).}

\ei

\subsection{Density correlation functions} \label{sec:g3}

The probability density of finding a spin-$\UP$ particle at~$\rr_1$ and a spin-$\down$ particle at $\rr_2$ is given by the pair correlation function
$g_2(\rr_1,\rr_2) = \la \hat{\psi}^\dagger_\UP(\rr_1) \hat{\psi}^\dagger_\down(\rr_2) \hat{\psi}_\down(\rr_2) \hat{\psi}_\UP(\rr_1) \ra = \la \hat{n}_\UP(\rr_1) \, \hat{n}_\down(\rr_2) \ra$.
Similarly, the probability density of finding a spin-$\UP$ particle at $\rr_1$,
a second spin-$\UP$ particle at $\rr_2$,
and a spin-$\down$ particle at $\rr_3$ is given by the triplet correlation function
\bea
g_{2,1}(\rr_1,\rr_2,\rr_3)
 &=& \la \hat{\psi}^\dagger_\UP(\rr_1) \hat{\psi}^\dagger_\UP(\rr_2) \hat{\psi}^\dagger_\down(\rr_3) \hat{\psi}_\down(\rr_3) \hat{\psi}_\UP(\rr_2) \hat{\psi}_\UP(\rr_1) \ra \nonumber
 \\&=&\la \hat{n}_\UP(\rr_1) \, \hat{n}_\UP(\rr_2) \, \hat{n}_\down(\rr_3) \ra
 \ -\ \delta^3(\rr_1-\rr_2)\ \la \hat{n}_\UP(\rr_1) \,\hat{n}_\down(\rr_3)\ra\,.
 \nonumber
 \eea
{\bl \\The second-order density correlation function has the short-distance asymptotic behavior~\cite{TanEnergetics}}
\be
\int d^3\!c\ \, g_2\!\!\left(\cc+\frac{\rr}{2},\cc-\frac{\rr}{2}\right) \ \underset{r\to0}{\sim}\ \, \frac{1}{(4\pi)^2}\ \, \frac{C_2}{r^2}\,. \nonumber
\ee
{\bl For the third-order density correlation function, we find the short-distance asymptotic behavior}
\be
\boxed{\int d^3\!C\ d^5\Omega\ \, g_{2,1}(\rr_1,\rr_2,\rr_3)\underset{R\to0}{\sim}\ 
C_{2,1}\,R^{2s-4}
\ \frac{32(s+1)}{3\sqrt{3}}}
\label{eq:g21_C3}
\ee
where the limit $R\to0$ is taken for fixed $\CC$ and~$\Oo$.
Here a change of coordinates is implied between $(\rr_1,\rr_2,\rr_3)$ and $(\CC,R,\Oo)$, 
with $R$ the hyperradius
defined in Eq.~(\ref{eq:def_R}),
$\CC = (\rr_1+\rr_2+\rr_3)/3$ the center-of-mass,
and
$\Oo$
 the hyperangles
which are five  dimensionless coordinates that remain to determine the positions of the three particles, see~Eq.~(\ref{eq:def_Oo}). 
{\bl Similarly, the $\UP\down\down$ triplet correlation function
$g_{1,2}(\rr_1,\rr_2,\rr_3)
= \la \hat{\psi}^\dagger_\UP(\rr_1) \hat{\psi}^\dagger_\down(\rr_2) \hat{\psi}^\dagger_\down(\rr_3) \hat{\psi}_\down(\rr_3) \hat{\psi}_\down(\rr_2) \hat{\psi}_\UP(\rr_1) \ra$
satisfies
\be
\boxed{\int d^3\!C\ d^5\Omega\ \, g_{1,2}(\rr_3, \rr_1,\rr_2)  \underset{R\to0}{\sim}\ 
C_{1,2}\,R^{2s-4}
\ \frac{32(s+1)}{3\sqrt{3}}.}
\label{eq:g12_C3}
\ee
}

\subsection{Link with the many-body wavefunction}
\label{sec:psi}

When three particles  approach each other,
the many-body wavefunction has a 
singular asymptotic behavior,
with a prefactor
related to three-body contact.
Let $\psi(\rr_1,\ldots,\rr_N)$ be the (orbital) many-body wavefunction. 
Without loss of generality,
we can assume $N_\UP\geq2$, 
and
consider that
particles $(1,2,3)$ have spins $(\UP,\UP,\down)$.
We take this convention throughout the article. 
This means that
$\psi$ is antisymmetric  w.r.t. exchange of $\rr_1$ and {\bl $\rr_2$}
($\psi$~is also antisymmetric w.r.t. exchange of any other pair of same-spin particles).
We take the normalization 
$\int \dr_1\ldots \dr_N\ |\psi(\rr_1,\ldots,\rr_N)|^2 = 1$.

Within the zero-range model, the stationary Schr\"odinger equation
\be
\sum_{i=1}^N \left[-\frac{\hbar^2}{2m}\Delta_{\rr_i} + U(\rr_i) \right] \psi = E\,\psi
\label{eq:schro}
\ee
contains an external trapping potential $U$
but no interaction potential;
instead, $\psi$ should satisfy a contact condition in the limit where two particles of opposite spin approach each other:
\\There exists $A$ such that
\be
\psi(\rr_1,\ldots,\rr_N)
\ \underset{r\to0}{=}\ 
\left(
\frac{1}{r} - \frac{1}{a_2}
\right)
\,A(\cc;\rr_2,\rr_4,\ldots,\rr_N) 
+ O(r)
\label{eq:BP}
\ee
where $a_2$ is the two-body scattering length,
$r = \| \rr_1 - \rr_3 \|$ is the distance between the opposite-spin particles 1 and~3, and $\cc = (\rr_1+\rr_3)/2$ is their center-of-mass.
The limit $r\to0$ is taken for fixed $\cc$ and fixed positions of the remaining particles $(\rr_2,\rr_4,\ldots,\rr_N)$.
By antisymmetry, a 
similar contact condition automatically also holds for all other pairs of opposite-spin particles, and Eqs.~(\ref{eq:schro},\ref{eq:BP}) are sufficient to define the eigenstates $\psi$ and energies $E$ of the zero-range model.\footnote{\bl Configurations with a vanishing interparticle distance are implicitly excluded in (\ref{eq:schro}). In an equivalent alternative formulation,  these configurations are included and regularized delta pseudopotential terms are added~\cite{Petrov3fermions,CRAS}.}

When particles 1, 2 and 3 approach each other, the wavefunction of any stationary state has the asymptotic behavior~\cite{Petrov4body2004,TanScalingREVTEX,LeChapitreIn2,Efimov}
\be
\boxed{\psi(\rr_1,\ldots,\rr_N)
\underset{R\to0}{\sim}
R^{s-2}\sum_{{\rm m}=-1}^{+1} \phi_{\rm m}(\Oo)\ B_\md(\CC;\rr_4,\ldots,\rr_N).}
\label{eq:R0}
\ee
Here, as in Eq.~(\ref{eq:g21_C3}), $R$ is the hyperradius of particles ($1$, $2$, $3$) defined in Eq.~(\ref{eq:def_R}),
$\CC$ is their center-of-mass,
and $\Oo$ denotes their
hyperangles.
The limit $R\to0$ is taken for fixed $\Oo, \CC,\rr_4,\ldots,\rr_N$.
The unitary hyperangular wavefunctions
$\phi_\md(\Oo)$ are
such that $R^{s-2}\,\phi_\md(\Oo)$ is a solution of the three-body problem at zero energy and infinite scattering length
with total angular momentum quantum numbers $l=1$ and $\md\in\{-1,0,1\}$,
see Appendix~\ref{app:unit_hyp} for more details.\footnote{\bl There is a similarity between the
two-body and three-body short-distance asymptotic behaviors  Eqs.~(\ref{eq:BP}) and~(\ref{eq:R0}),
given that $1/r - 1/a_2$ is a solution of the
two-body problem at zero energy.}$^{,}$\footnote{\bl An asymptotic behavior  similar to (\ref{eq:R0}) holds when any three  particles with 
spins $\UP\UP\down$ approach each other,
the  functions $B_\md$ corresponding to different triplets of particles 
being simply related to each other by antisymmtery.}

It is known~\cite{TanEnergetics} that $C_2$ is given by the norm of the function $A$ that appears in Eq.~(\ref{eq:BP}),
\be
C_2 = (4\pi)^2\,
N_\UP\,N_\down
\int \left|A(\cc;\rr_2,\rr_4,\ldots,\rr_N)\right|^2 \,d^3\!c\,\dr_2\,\dr_4\ldots \dr_N.
\label{eq:C2_AA}
\ee
Similarly, 
the three-body contact
is given by the norm of the function $B$ that appears in Eq.~(\ref{eq:R0}),
\be
\boxed{
C_{2,1} = 
N_\UP (N_\UP-1) N_\down
\ \frac{3\sqrt{3}}{32\,(s+1)}\ 
\\ \sum_{\md=-1}^{+1}
\int \left|B_\md(\CC;\rr_4,\ldots,\rr_N)\right|^2
d^3\!C\, \dr_4 \ldots \dr_N.}
\label{eq:C3_BB}
\ee
The expression (\ref{eq:N21}) for the number of nearby $\UP\UP\down$ fermion triplets,
together with Eq.~(\ref{eq:C3_BB}),
simply follow from Eq.~(\ref{eq:R0})
by integrating $|\psi|^2$ over the $R<\epsilon$ region.\footnote{\bl Here we used the change of integration variables
$(\rr_1,\rr_2,\rr_3)\longrightarrow(\rr,\rrho,\CC)$
with $\rrho$ defined in (\ref{eq:def_jaco}),
 of Jacobian $|\partial(\rr_1,\rr_2,\rr_3)/\partial(\rr,\rrho,\CC)| = (\sqrt{3}/2)^3$.
We also used the property $\int d^5\Omega\ \phi_\md(\Oo)^*\,\phi_{\md'}(\Oo) = \delta_{\md,\md'}$.}
Similarly, the relation involving $g_{2,1}$, Eq.~(\ref{eq:g21_C3}), 
follows immediately from Eqs.~(\ref{eq:R0},\ref{eq:C3_BB}) {\gre and
from the expression of $g_{2,1}$ in first quantization,
$g_{2,1}(\rr_1,\rr_2,\rr_3)  =  
 N_\UP(N_\UP-1)N_\down
\int \dr_4 \ldots \dr_N\,|\psi(\rr_1,\ldots,\rr_N)|^2$.}

There is a completely analogous relation between $C_{1,2}$ and the behavior of the many-body wavefunction when three particles of spins $\UP\down\down$ approach each other (provided $N_\down \geq 2$).
Specifically, considering that particle 4 has spin $\down$
(while particles 1, 3 still have spins $\UP, \down$)
and denoting by $\tilde{R}$, $\tilde{\Oo}$, $\tilde{\CC}$ the hyperradius, hyperangles and center-of-mass associated to particles $3,4,1$
[obtained by replacing $(\rr_1,\rr_2,\rr_3)$ with
$(\rr_3,\rr_4,\rr_1)$ in
 Eqs.~(\ref{eq:def_jaco},\ref{eq:def_RR},\ref{eq:def_Oo})]
we have
\be
\boxed{\psi(\rr_1,\ldots,\rr_N)
\underset{\tilde{R}\to0}{\sim}
\tilde{R}^{s-2}\!\!\sum_{{\rm m}=-1}^{+1}\!\! \phi_{\rm m}(\tilde{\Oo})\ \tilde{B}_\md(\tilde{\CC};\rr_2,\rr_5,\ldots,\rr_N)}
\label{eq:R0t}
\ee
which yields the relations for $N_{1,2}$ and $g_{1,2}$, Eqs.~(\ref{eq:N12},\ref{eq:g12_C3}), with
\be
\boxed{C_{1,2} = 
N_\down (N_\down-1) N_\UP
\ \frac{3\sqrt{3}}{32\,(s+1)}
\ \, \sum_{\md=-1}^{+1}
\ \, \int \left|\tilde{B}_\md(\CC;\rr_2,\rr_5,\ldots,\rr_N)\right|^2
d^3\!C\, \dr_2\,\dr_5 \ldots \dr_N.}
\label{eq:C3_BBt}
\ee
Here we assumed $N_\down\geq2$;
if $N_\down=1$, then $N_{1,2}$ is obviously zero,
and $C_{1,2}=0$.

\vskip 0.1cm
\noindent{\bl \underline{\it Remark:}\ \ Higher-body contacts can be defined in the same way than the three-body contacts.
When $j_\UP$ particles of spin $\UP$ and $j_\down$ particles of spin $\down$ approach each other, the $N$-body wavefunction factorizes into the product of {\it (i)} a function of the relative positions of the $j=j_\UP+j_\down$ nearby particles, given by a zero-energy solution of the $j_\UP+j_\down$ body problem at $a_2{=}\infty$, proportional to the $j$ body hyperradius to some power, and {\it (ii)} a function of the center-of-mass of the $j$ nearby particles and of the positions of the $(N{-}j)$ other particles~\cite{TanScalingREVTEX,LeChapitreIn2}.
The $L^2$ norm of the latter function defines the $j_\UP+j_\down$ body contact $C_{j_\UP,j_\down}$ (up to a prefactor which is a matter of definition).}


\subsection{Large-momentum tail {\bl of the center-of-mass momentum distribution of nearby pairs}}
\label{sec:tail_NP}

Since $C_2$ and $C_3$ determine short-distance singularities, it is natural that they also determine large-momentum tails.
$C_2$ determines the leading tail of 
the single-particle momentum distribution~\cite{TanLargeMomentum}
\be
N_\sigma(\kk) \ \underset{k\to\infty}{\sim} \ \frac{C_2}{k^4}
\label{eq:nk_C2}
\ee
with the normalization $\int N_\sigma(\kk)\, d^3k/(2\pi)^3 = N_\sigma$
(in the case of periodic boundary conditions, momenta become discrete and momentum integrals should be replaced by sums).

$C_3$ also determines a 
large-momentum tail. Suppose that one measures, for a pair of particles with opposite spin, both their spatial separation $r$ and their center-of-mass momentum $\KK$
(this is allowed 
since the corresponding operators commute).
Let $N_2(\epsilon, \KK)$ be the probability distribution over $\KK$
conditional to $r<\epsilon$,
with the normalization 
$\int N_2(\epsilon,\KK)\, \dK/(2\pi)^3 = N_2(\epsilon)$.
{\bl In other words, $N_2(\epsilon, \KK)$ is the center-of-mass momentum distribution of the pairs of particles separated by a distance $<\epsilon$, normalized to the total number of such pairs.\footnote{\gre More formally, $N_2(\epsilon, \KK)$ is the expectation value of the operator $\ds \sum_{i:\UP, j:\down} \theta(\epsilon-\hat{r}_{ij})\ (2\pi)^3\,\delta^3(\hat{\KK}_{ij}-\KK)$, where $\theta$ is the Heaviside function,
while $\hat{r}_{ij} = \| \hat{\rr}_j - \hat{\rr}_i \|$ and $\hat{\KK}_{ij}=\hat{\kk}_i + \hat{\kk}_j$ are the operators corresponding to the relative distance and the center-of-mass momentum of particles $i$ and $j$.}}
We have
\be
N_2(\epsilon, \KK) \ \underset{\epsilon\to0}{\sim}\ \, N_P(\KK)\ \frac{\epsilon}{4\pi}
\label{eq:def_NP}
\ee
where $N_P(\KK)$ 
{\bl is what we call the}
 center-of-mass momentum distribution of  
nearby fermion pairs.
{\bl With this definition, one simply has the normalization}
\be \int N_P(\KK)\,\frac{d^3\!K}{(2\pi)^3} = C_2 \label{eq:int_NP} \ee
{\bl as a consequence of~(\ref{eq:N2}).}\footnote{$N_P(\KK)$ appears naturally in the diagrammatic formalism:
In a homogeneous system, $N_P(\KK)$ divided by the volume equals $-(m^2/\hbar^2)\,\Gamma(\KK,\tau=0^-)$ where $\Gamma$ is the pair-propagator defined {\it e.g.} in~\cite{RossiContact}. This can be shown using a lattice model~\cite{WernerCastinRelationsFermions}, for which $1/(r\,r')$ in (\ref{eq:rho2_BP}) can be replaced by $\phi(\rr)\phi(\rr')$ with $\phi$ the zero-energy two-body scattering state, and setting $\rr=\rr'=\vn$.}
The tail of $N_P(\KK)$ is determined by the three-body contact\footnote{The fact that $\Gamma(K,\tau=0^-)$ has a tail $\propto C_3 / K^{2s+4}$ was pointed out to us by Shina Tan (private communication, Aspen, 2011).}:
\be
  \boxed{
    \bar{N}_P(K) \ \underset{K\to\infty}{\sim} \ \ {\bl \Mr}\ \frac{C_3}{K^{2s+4}}}
\label{eq:NP_tail}
  \ee
with the prefactor
\be
 \boxed{{\bl \mathcal{M}} = 32\, \pi^3\,\frac{4^s}{3^{s+1/2}}\,(s+1)\,\Gamma(s+2)^2\,\sin^2(s\pi)\,\Nr^2}
\label{eq:Mr}
\ee
whose numerical value is
\be
\boxed{\Mr \simeq 2272.}
\label{eq:Mr_val}
\ee
Here $\bar{N}_P(K)$ stands for the angular average  $\int N_P(\KK)\,d{\bf \hat{K}}/(4\pi)$ {\bl (with ${\bf \hat{K}} := \mathbf{K} / K$, 
and $d{\bf \hat{K}}$ the differential solid angle, so that $d^3\!K = d{\bf \hat{K}}\ K^2\,d\!K$).}

To derive this result, we
consider the two-body reduced density matrix
$\rho_2(\rr_\UP, \rr_\down; \rr'_\UP, \rr'_\down) := \la \hat{\psi}^\dagger_\UP(\rr'_\UP) \hat{\psi}^\dagger_\down(\rr'_\down) \hat{\psi}_\down(\rr_\down) \hat{\psi}_\UP(\rr_\UP) \ra$
or equivalently in first quantization
\be
\rho_2(\rr_\UP, \rr_\down; \rr'_\UP, \rr'_\down) = N_\UP N_\down \int \dr_2\,\dr_4 \ldots \dr_N
\
\psi^*\!(\rr'_\UP, \rr_2, \rr'_\down,\rr_4, \ldots , \rr_N)
\ \psi(\rr_\UP, \rr_2, \rr_\down,\rr_4, \ldots , \rr_N).
\label{eq:rho2_1q}
\ee
Inserting the two-body contact condition (\ref{eq:BP}) into (\ref{eq:rho2_1q}) 
 yields
\be
\rho_2\left( \cc - \frac{\rr}{2}\,,\, \cc + \frac{\rr}{2} \,;\,
\cc'-\frac{\rr'}{2}\,,\, \cc'+\frac{\rr'}{2} \right)
\ \underset{r,r'\to0}{\sim}
\ \ \frac{g_P(\cc,\cc')}{(4\pi)^2\, r\, r'}
\label{eq:rho2_BP}
\ee
where
\be
g_P(\cc,\cc') = (4\pi)^2\ N_\UP N_\down \int \dr_2\, \dr_4 \ldots \dr_N
\ A^*\!(\cc'; \rr_2, \rr_4, \ldots, \rr_N)
\ A(\cc; \rr_2, \rr_4, \ldots, \rr_N)\,.
\label{eq:P_vs_A}
\ee
We see that $g_P$ can be physically interpreted as a coherence function for pairs of nearby fermions.
Accordingly, $N_P$ is related to $g_P$ by Fourier transformation,
\be
N_P(\KK) = \int \dc\,\dc'\ \,e^{-i\, \KK \cdot (\cc - \cc')} \ g_P(\cc,\cc')
\label{eq:NP_gP}
\ee
{\bl as can be formally shown using the definition (\ref{eq:def_NP}) of $N_P$.} Hence
\be
  N_P(\KK) = (4\pi)^2\,N_\UP\,N_\down\,\int \dr_2\,\dr_4\ldots \dr_N
  \\ \left| \int \dc\ e^{-i\, \KK \cdot \cc}\ A(\cc; \rr_2, \rr_4, \ldots, \rr_N) \right|^2.
  \label{eq:N_P_AA}
\ee
We then follow a reasoning resembling  the one used
to derive Eq.~(\ref{eq:nk_C2})
in Sec.~IV.A of~\cite{WernerCastinRelationsFermions}.
\\In the large $K$ limit, the Fourier transform with respect to $\cc$ in (\ref{eq:N_P_AA}) is dominated by the contributions from the singularities of
$A(\cc; \rr_2, \rr_4, \ldots, \rr_N)$, which occur when $\cc$ approaches one of the $\rr_i$ ($i=2, 4, \ldots, N$).
This corresponds to 
particles 1, 3 and~$i$ being close to each other
[since $A(\cc; \rr_2, \ldots)$ determines the wavefunction $\psi$ when particles 1 and 3 are close to $\cc$, according to Eq.~(\ref{eq:BP})].
For example, for $i=2$, the behavior of $A(\cc; \rr_2, \rr_4, \ldots, \rr_N)$ in the limit $\cc\to\rr_2$ is determined by the asymptotic behavior of $\psi$ when the three particles 1, 2, 3 are close, given by Eq.~(\ref{eq:R0}).
Therefore, we just need to
take the limit where particles 1 and 3 approach each other in (\ref{eq:R0}) to obtain
\be
A(\rr_2-\uu; \rr_2, \rr_4, \ldots, \rr_N)
\underset{u\to0}{\sim}\ \ 
\Nr\ \frac{s}{2}\ \cos\left(\frac{s\pi}{2}\right)
\left( \frac{2\,u}{\sqrt{3}} \right)^{s-1}
\, \sum_{\md=-1}^1
\,Y_1^\md({\bf \hat{u}})
\, B_\md(\rr_2; \rr_4, \ldots , \rr_N)
\nonumber
\ee
where we used the expression (\ref{eq:phi},\ref{eq:varphi_l=1}) of $\phi_\md(\Oo)$.
There is a similar singularity of $A(\cc; \rr_2, \rr_4, \ldots, \rr_N)$ when $\cc$ approaches $\rr_i$ with $4\leq i\leq N$;
when particle $i$ has spin $\down$, the function $\tilde{B}_\md$ introduced in~(\ref{eq:R0t}) appears instead of $B_\md$.
This gives
\begin{multline}
\int \dc\ e^{-i\, \KK \cdot \cc}\ A(\cc; \rr_2, \rr_4, \ldots, \rr_N)
\underset{K\to\infty}{\sim} 
\ \  \frac{2^{s-2}}{3^{(s-1)/2}}
\ \Nr\, \cos\left(\frac{s\pi}{2}\right)
\sum_{\md=-1}^1
I_\md(\KK) 
\\ \times \left[
  \sum_{i:\UP, i\neq1}
  (1-2\,\delta_{i,2})\,e^{-i\KK\cdot\rr_i}
  \,  B_\md\left(
P_{2i}(\rr_2;\rr_4,\ldots,\rr_N) \right)
\right.
   \left. +  \sum_{i:\down, i\neq3}
  (1-2\,\delta_{i,4})\,e^{-i\KK\cdot\rr_i}
   \,  \tilde{B}_\md\left(
P_{4i}(\rr_2;\rr_4,\ldots,\rr_N)\right)
  \right]
\label{eq:TF_A}
\end{multline}
where
the first (resp. second) sum over $i$ 
is taken over particles with spin $\UP$ (resp. $\down$), 
$P_{ji}(\rr_2;\rr_4,\ldots,\rr_N)$ is obtained from $(\rr_2;\rr_4,\ldots,\rr_N)$ by exchanging $\rr_j$ with $\rr_i$ ($P_{ii}$ is the identity),
and
$I_\md(\KK) := s \int d^3\!u\, e^{i\KK\cdot\uu}\, u^{s-1}\, 
Y_1^\md({\bf \hat{u}})$.
The latter integral  can be evaluated analytically:
Using $\int d{\bf \hat{u}}\ e^{i \KK\cdot\uu}\, Y_1^\md({\bf \hat{u}}) = 4\pi i \  Y_1^\md(\hat{K})\, j_1(K u)$ with $j_1(t) = (\sin t  - t \cos t) / t^2$,
and evaluating the remaining integral over $u$ by integrating along a closed contour including the positive real axis and negative imaginary axis, we get
\be
I_\md(\KK) = 4\pi i\ \Gamma(s+2)\,\sin\left(\frac{s\pi}{2}\right)\ \frac{Y_1^\md({\bf \hat{K}})}{K^{s+2}}.
\label{eq:I_K}
\ee
Inserting (\ref{eq:TF_A},\ref{eq:I_K}) into (\ref{eq:N_P_AA}), expanding the modulus squared,
and neglecting in the large $K$ limit
the cross terms coming from two different values of $i$,\footnote{By power counting, these cross terms
{\gre give rise to a $\propto 1/K^{2 s_4 + 4}$ tail of $\bar{N}_P(K)$,}
  where $s_4$ is the smallest scaling exponent of the unitary four-fermion problem;
this is indeed negligible compared to the leading $1/K^{2 s+{\gre 4}}$ tail of $\bar{N}_P(K)$,
given that $s_4 = 2.509(1)$ is larger than $s \equiv s_3$. This value of $s_4$ follows from the four-body ground-state energy $E_4$ in an isotropic harmonic trap computed in~\cite{DailyBlume_4body_spectrum} and the relation~\cite{TanScalingREVTEX,WernerSym} $E_N = (s_N + 5/2) \hbar\omega$.}
we obtain the result~(\ref{eq:NP_tail},\ref{eq:Mr},\ref{eq:Mr_val}), {\bl where
we used the value of $\Nr$ given in Eq.~(\ref{eq:Nr}) of Appendix~\ref{app:unit_hyp}.}

{\gre

  \subsection{Large-momentum tail of the two-particle momentum distribution}
  \label{sec:N_k1k2}

The three-body contact also determines the asymptotic behavior at large momenta of the two-particle opposite-spin momentum distribution function, defined by
  \be
  N(\kk_1,\kk_2) = \left\langle \hat{N}_\UP(\kk_1)\,\hat{N}_\down(\kk_2) \right\rangle
\label{eq:def_Nk1k2}
  \ee
  where $\hat{N}_\sigma(\kk) := \hat{c}^\dagger_\sigma(\kk)\,\hat{c}_\sigma(\kk)$
  with $\hat{c}_\sigma(\kk) = \int \dr \ e^{- i \kk \cdot \rr}\ \hat{\psi}_\sigma(\rr)$ the annihilation operator of a particle of spin $\sigma$ in the state $|\kk\ra$ defined by $\la \rr | \kk \ra = e^{i\kk\cdot\rr}$.
\\Since $\int \hat{N}_\sigma(\kk) \, \frac{d^3k}{(2\pi)^3} = N_\sigma$, we have the normalization $\int N(\kk_1,\kk_2) \, \frac{d^3k_1}{(2\pi)^3}\, \frac{d^3k_2}{(2\pi)^3} = N_\UP\,N_\down$.
\\Experimentally, the two-particle momentum distribution 
can be accessed from the statistics of time-of-flight images, as was recently demonstrated in 2D~\cite{Jochim_nk_noise}.\footnote{In~3D, early measurements in the BEC regime were reported in \cite{Jin_nk_noise}; see also \cite{Braun_n_k1_k2} for a recent numerical study.
In optical lattices,
detailed experimental studies were carried out in recent years using metastable helium atoms~\cite{Clement_HBT,Clement_pairs_opposite_k,Clement_FCS}.}

Taking the limit of a large relative momentum $k\to\infty$, and integrating over the center-of-mass momentum $\KK$, one obtains a tail
proportional
to $C_2$\,,
  \be
  \int N\left(\frac{{\bf K}}{2}-{\bf k}\,,\,\frac{{\bf K}}{2}+{\bf k}\right)\ \frac{\dK}{(2\pi)^3}\ \ \underset{k\to\infty}{\sim}\ \ \frac{C_2}{k^4}
  \label{eq:Nkk_C2}
  \ee
  as pointed out in \cite{BarneaFewBodySys}.
  If we instead send 
  $K$ to infinity, and average over the direction of $\KK$, we obtain a tail proportional to $C_3$\,,
  \be
  \boxed{\lim_{K\to\infty}\ \lim_{k\to\infty} \ K^{2s+4}\ k^4\ \frac{1}{4\pi} \int d{\bf \hat{K}}\ \ N\left(\frac{{\bf K}}{2}-{\bf k}\,,\,\frac{{\bf K}}{2}+{\bf k}\right)\ \ = \ \Mr\ C_3}
  \ee
  where the constant $\Mr$ was given in (\ref{eq:Mr}).

This result immediately follows from (\ref{eq:NP_tail}) and from the relation
\be
N\left(\frac{{\bf K}}{2}-{\bf k}\,,\,\frac{{\bf K}}{2}+{\bf k}\right) \ \ \underset{k\to\infty}{\sim}\ \ \frac{N_P(\KK)}{k^4}\,,
\label{eq:Nkk_NP}
\ee
obtained in~\cite{BarneaFewBodySys} and rederived in the sequel.\footnote{Relation~(\ref{eq:Nkk_C2}) also follows from~(\ref{eq:Nkk_NP}), given (\ref{eq:int_NP}).}$^{,}$\footnote{In Ref.~\cite{BarneaFewBodySys}, $N_P(\KK)$ was defined by
  \be
  N_P(\KK) = (4\pi)^2\ \int \ \frac{d^3k_2}{(2\pi)^3}\ \frac{d^3k_4}{(2\pi)^3} \ \ldots \ \frac{d^3k_N}{(2\pi)^3}\ \left| \tilde{A}(\KK; \kk_2, \kk_4 , \ \ldots \ ,\kk_N) \right|^2
\nonumber
  \ee
  with $\tilde{A}$ the Fourier transform of $A$
  $$\tilde{A}(\KK; \kk_2, \kk_4 , \ \ldots \ ,\kk_N) \ := \ \int \dc \ \dr_2\ \dr_4\,\ldots \, \dr_N\ e^{-i ( \KK \cdot \cc + \kk_2 \cdot \rr_2 + \kk_4 \cdot \rr_4 + \ldots + \kk_N \cdot \rr_N ) }\ A(\cc ; \rr_2, \rr_4 , \, \ldots \, , \rr_N)\,,$$
  and~(\ref{eq:Nkk_NP}) was deduced from the expression
  $$N(\kk_1,\kk_3) = N_\UP\,N_\down\ \int \ \frac{d^3k_2}{(2\pi)^3}\ \frac{d^3k_4}{(2\pi)^3} \ \ldots \ \frac{d^3k_N}{(2\pi)^3}\ \left| \tilde{\psi}(\kk_1, \, \ldots \, , \kk_N) \right|^2$$
  in terms of the momentum-space wavefunction
  $$\tilde{\psi}(\kk_1, \, \ldots \, , \kk_N) \, := \, \int\,\dr_1 \, \ldots \dr_N\ e^{-i (\kk_1 \cdot \rr_1 + \ldots + \kk_N \cdot \rr_N)}\,\psi(\rr_1 , \, \ldots \, , \rr_N)$$
which has the asymptotic behavior
  $$\tilde{\psi}\left(\frac{\KK}{2}-\kk, \, \kk_2, \, \frac{\KK}{2}+\kk, \, \kk_4, \  \ldots \  , \kk_N\right)\ \underset{k\to\infty}{\sim}\ \frac{4\pi}{k^2}\ \tilde{A}(\KK; \kk_2, \kk_4 , \ \ldots \ , \kk_N)\,.$$
} 
The definition (\ref{eq:def_Nk1k2}) yields
\begin{equation}
N\left(\frac{{\bf K}}{2}-{\bf k}\,,\,\frac{{\bf K}}{2}+{\bf k}\right)=
\int \dc\, \dc' \,\dr\, \dr'
e^{i{\bf K}\cdot({\bf c}'-{\bf c})}e^{i{\bf k}\cdot({\bf r}'-{\bf r})}
\ \rho_2\!\left({\bf c}-\frac{{\bf r}}{2},{\bf c}+\frac{{\bf r}}{2};{\bf c}'-\frac{{\bf r}'}{2},{\bf c}'+\frac{{\bf r}'}{2}\right)
\label{eq:N2_rho2}
\end{equation}
where $\rho_2$ is the two-body reduced density matrix, which has the diverging behavior~(\ref{eq:rho2_BP}) when $r$ and $r'$ tend to zero.
This short-distance divergence leads to a $k\to\infty$ tail of the Fourier transform Eq.~(\ref{eq:N2_rho2}), which can be computed by replacing $\rho_2$ with its asymptotic expression~(\ref{eq:rho2_BP}), and using the identity (in the sense of distributions) $\int \dr \, e^{-i\kk\cdot\rr}/r = 4\pi/k^2$.
Using (\ref{eq:NP_gP}) then yields (\ref{eq:Nkk_NP}).
  
}

{\bl
 
\section{Relations for finite-range models} \label{sec:finite_b}

In this Section,
we go beyond the zero-range model, and consider interactions of small but non-zero range.
We express two observable in terms of the three-body contact: the rate of three-body recombinations towards deeply bound dimers (Sec.~\ref{sec:loss}), and the three-body contribution to the energy difference between finite-range and zero-range models (Sec.~\ref{sec:dE}).


}

\subsection{Three-body loss rate} \label{sec:loss}

 In ultracold atom experiments, three-body losses generically take place,
being a manifestation of the fact that the true equilibrium state at such low temperatures is not gaseous
(with the exception of polarized hydrogen).
In this Section we relate the rate of three-body losses to the three-body contact.

\subsubsection{Simple finite-range interaction} \label{subsec:simple}

To describe three-body losses, we need to go beyond the zero-range model.
In this subsection we consider a simple model where fermions of different spin interact through a {\bl rotationally invariant} potential $V_2(r)$, of finite range $b$. \footnote{\bl Typically, $b$ is set by the true range $b_0$ of the potential ({\it i.e.} the length such that $V_2(r)$ decays quickly for $r\gg b_0$). More generally, $b = {\rm Max}(b_0, |r_e|)$ where $r_e$ is the effective range.}
We consider the resonant regime
where the two-body scattering length $a_2$ is large,
\be
|a_2|\gg b.
\label{eq:a>>b}
\ee
In this regime, there are two kinds of two-body bound states:
\begin{itemize}
\item
  the weakly bound dimer, of binding energy $\approx \hbar^2/(m a_2^{\phantom{2}2})$, which exists for $a_2>0$
\item deeply bound dimer(s), of binding energy ${\bl \gtrsim} \, \hbar^2/(m b^2)$,
  which exist if the interaction potential is deep enough {\bl (as in generic cold atom experiments)}.
     \end{itemize}
We consider a stationary solution of the $N$-body Schr\"odinger equation
\be
H\ \psi = E\ \psi
\label{eq:schro_V2}
\ee
\be
H = \sum_{i=1}^N \left[-\frac{\hbar^2}{2m}\Delta_{\rr_i} + U(\rr_i) \right] 
\ +\  \sum_{\substack{i:\UP,\,j:\down}} V_2(r_{ij})
\label{eq:H_V2}
\ee
in the zero-range regime
\be
1/\ktyp \gg b
\label{eq:ZRreg}
\ee
where $1/\ktyp$ is defined as the smallest scale of variation of the stationary wavefunction $\psi$ in the region where all interparticle distances are $\gg b$. \footnote{For example, for the ground state of the homogeneous unpolarized gas,
$\ktyp$ is $\sim k_F$ for $a_2<0$ and $\sim{\rm Max}\left(k_F, 1/a_2 \right)$ for $a_2>0$,
where $k_F$ is the Fermi momentum;
for the ground state of a few particles in an isotropic harmonic trap of frequency $\omega$,
$1/\ktyp$ is $\sim a_{\rm ho}$ for $a_2<0$ and $\sim {\rm Min}(a_{\rm ho}, a_2)$ for $a_2>0$,
where $a_{\rm ho}:=\sqrt{\hbar/(m\omega)}$ is the harmonic oscillator length.}


Let us first consider the case where there are no deeply bound states.
{\bl For simplicity we also assume that the spectrum is discrete (which is the case in a trapping potential of infinite depth --{\it e.g.} a harmonic trap-- or in a box with periodic boundary conditions).}
Then, 
in the zero-range regime (\ref{eq:ZRreg}),
the zero-range model is valid for any stationary state $\psi$,
in the sense that standard observables
tend to their respective values within the zero-range model.
This includes observables such as the energy, 
as well as 
$N_2(\epsilon)$ and $N_3(\epsilon)$
provided $\epsilon\gg b$
[to reach the asymptotic regimes of Eqs.~(\ref{eq:N2},\ref{eq:N3}) one also needs $\epsilon \ll 1/\ktyp$].

We turn to the experimentally relevant case where deeply bound dimers exist.
These deeply bound dimers can be formed through recombination processes between three atoms.
Let us denote by  $\Gamma_3$ the number of such events per unit of time.
The recombination products {\bl (the deeply bound dimer and the third atom)} escape from the trapped gas,
provided the trapping potential $U(\rr)$ has a finite depth much smaller than the binding energy ($\sim \frac{\hbar^2}{m b^2}$) of deeply bound dimers.
{\bl In typical experiments, this condition holds, and other loss processes are negligible, which allows one to measure $\Gamma_3$ from the decay of the number of trapped atoms, $\dot{N} = - 3\,\Gamma_3$~\cite{JinLifetime,SalomonCrossover,PetrovJPhysB,ThomasLosses,LuoL3Narrow}.}

In the zero-range regime, 
{\bl this decay is slow} compared to the other timescales of the problem,
as we will see.
A standard way to describe such a slowly decaying state in quantum mechanics is to consider a quasi-stationary Gamow state, {\it i.e.},
a solution of the Schr\"odinger equation 
with a complex energy
and an outgoing-wave asymptotic behavior~{\bl\cite{Gamow,LandauLifschitzMecaQnote,Messiah_Vol1_gamow,Taylor_Scattering_gamow,BlattWeisskopf,MichelGamowShellModel}}.
Accordingly, we will consider a solution $\psi$ of 
(\ref{eq:schro_V2},\ref{eq:H_V2})
with a complex $E$,
and an outgoing-wave asymptotic behavior corresponding to the recombination products (a deeply bound dimer + an atom) flying apart towards large distances.\footnote{\bl An alternative approach would be to use the Lindblad equation.
We expect that this would lead to the same result for the loss rate, as was checked for three-body losses for bosons in~\cite{BraatenLindblad}.} 
For such a quasi-stationary state,
in the zero-range regime,
standard observables again
tend to their respective values within the zero-range model.\footnote{
{\bl An appropriate normalization of the Gamow state will be given below in Eq.~(\ref{eq:norm}). Similarly, the expectation value of an observable $\hat{O}$ should be defined as $\int_\Rr d^{3N}\!X \ \int_\Rr d^{3N}\!X' \ \, \psi^*\!(\XX)\ \la\XX|\hat{O}|\XX'\ra\ \psi(\XX')$.}}$^{,}$\footnote{\bl Within the zero-range model, it is convenient to add steep infinite walls to the trapping potential at the boundary of $\Rr_{\rm trap}$, in order to have truly stationary states, thereby neglecting the exponentially suppressed evaporation process discussed in footnote~\ref{fn:evap}.}
The three-body loss rate $\Gamma_3$, however, is simply zero within the zero-range model. To compute $\Gamma_3$, one thus needs to go beyond the zero-range model. As we will see, one only needs to do so for the three-body problem,
in order to define a three-body parameter $a_3$. We then find
\be
\boxed{\Gamma_3 \simeq -\frac{\hbar}{m}\ 8 s (s+1) \ C_3\ {\rm Im} \, a_3 }
\label{eq:Gamma3}
\ee
where $C_3$ can be evaluated within the zero-range model.
Furthermore, breaking up $\Gamma_3$ into the sum of the two contributions $\Gamma_{2,1}$ and $\Gamma_{1,2}$ corresponding to $\UP\UP\down$ and $\UP\down\down$ loss processes, we have
\be
\boxed{\Gamma_{2,1} \simeq -\frac{\hbar}{m}\ 8 s (s+1) \, C_{2,1}\ {\rm Im} \, a_3}
\label{eq:Gamma2,1}
\ee
\be
\boxed{\Gamma_{1,2} \simeq -\frac{\hbar}{m}\ 8 s (s+1) \, C_{1,2}\ {\rm Im} \, a_3\,.}
\label{eq:Gamma1,2}
\ee
We expect these relations to be asymptotically exact in the resonant zero-range regime (\ref{eq:a>>b},\ref{eq:ZRreg}).

To define the three-body parameter $a_3$, we consider the zero-energy
{\bl free-space}
solution of the 
Schr\"odinger equation (\ref{eq:schro_V2},\ref{eq:H_V2})
for three particles
of spins $\UP\UP\down$
and angular-momentum quantum numbers $(l=1,\md)$
whose asymptotic behavior 
has the form
\be
\boxed{\Psi_\md(\RR) 
\simeq
\left(
R^s - \frac{a_3}{R^s}
\right)
\frac{1}{R^2}\ 
\phi_\md(\Oo)}
\label{eq:def_a3}
\ee
in the region $\{ b \ll r_{ij} \ll |a_2|, \ \forall i<j \} $
where all interparticle distances are large compared to the range 
but small compared to the two-body scattering length.
Here we have neglected the {\it deep-dimer + atom} outgoing wave, since it is proportional to the dimer wavefunction which is exponentially suppressed at distances $\gg b$.
The fact that $a_3$ does not depend on the quantum number~$\md$ follows from rotational invariance of the interaction.

\vskip 0.12cm
\noindent \underline{\it Remarks:}
\bi
\item
  {\bl Relation (\ref{eq:Gamma3}) is 
    reminiscent of the known relation~\cite{BraatenC,BraatenLindblad} between two-body loss rate and two-body contact~\footnote{
    Relation (\ref{eq:Gamma2}) concerns the 
    situation where the states $\UP$ and~$\down$ populated in the gas are not the two energetically lowest atomic internal states, 
    so that inelastic two-body collisions towards 
    lower lying states
are energetically allowed,
and $a_2$~acquires an imaginary part.}$^,$\footnote{\label{fn:Gamma2}
    \brown Relation (\ref{eq:Gamma2}) was obtained in~\cite{BraatenC} using the relation (\ref{eq:dE_da2}) and ${\rm Im}\,E = -\hbar\,\Gamma_2 /2$. 
    It was rederived in~\cite{BraatenLindblad} using the Lindblad equation.
    For completeness, we note that the relation~(\ref{eq:Gamma2}) can also be derived by a flux computation.
    The main steps of this derivation are as follows. We consider a solution $\psi$ of the zero-range model (\ref{eq:schro},\ref{eq:BP}) with complex values of $1/a_2$ and $E$. The corresponding time-dependent wavefunction is $\Psi(t) = \psi\,e^{-i E t / \hbar}$, and we have $\Gamma_2 = - \partial_t\la\Psi(t)|\Psi(t)\ra|_{t=0}$. Using (\ref{eq:continu},\ref{eq:current}), we obtain that $\Gamma_2$ equals $N_\UP N_\down$ times the limit when $\epsilon\to0$ of the probability flux entering into the region $\{ (\rr_1, \ldots , \rr_N) | r < \epsilon \}$, which can be simplified to
    $\Gamma_2 = - 2 N_\UP N_\down \frac{\hbar}{m} \lim_{\epsilon\to0} \int \dc \, \dr_2 \, \dr_4 \ldots \dr_N \, \epsilon^2 \int d{\bf \hat{r}} \ {\rm Im}(\psi^* \partial_r \psi)|_{r=\epsilon}$. Using (\ref{eq:BP}) then yields (\ref{eq:Gamma2}).}
\be
\Gamma_2 = \,\frac{\hbar}{2\pi m}\ C_2\ \,{\rm Im}\!\left(\frac{1}{a_2}\right).
\label{eq:Gamma2}
\ee }

\item
  {\bl ${\rm Im} \, a_3$ must be negative, since the loss rate is positive.}

\item
  {\bl Typically}
  one has the order of magnitude estimate $a_3 \sim b^{2s}$,
  assuming that there is no
 \\ {\bl extra} fine-tuning.\footnote{Indeed, the behavior (\ref{eq:def_a3}) has to be matched at $R$ of order $b$
with the solution inside the potential,
which {\bl typically} imposes that the two terms in (\ref{eq:def_a3}) are of the same order of magnitude for $R\sim b$~\cite{Petrov3fermions,PetrovLesHouches2010}.
For simplicity, we excluded here the special regime $|a_3| \gg b^{2s}$ that corresponds to the vicinity of a three-body resonance
(see~\cite{SonNbodyResDecay,PricoupenkoNbodyRes}, and~\cite{LeChapitreIn2,WernerSym,NishidaSonTan3bRes,Blume3bodyResPRL,GandolfiCarlson3bRes,Blume3bodyResPRA,Sadeghpour_Blume_3bodyRes,Kartavtsev3bRes} for the mass-imbalanced case with $0\leq s < 1$).
  Reaching this regime  would require a second fine-tuning of the interaction, in addition to the first fine-tuning that causes $|a_2| \gg b$ (for cold atoms, it would require a second control parameter of the interaction, in addition to the magnetic field used to tune $a_2$ to large values).
  {\bl We expect the relations (\ref{eq:Gamma3},\ref{eq:Gamma2,1},\ref{eq:Gamma1,2}) and the other results of this article to remain applicable in the three-body resonant regime provided the three-body parameter(s) remain(s) small (in modulus) compared to $1/\ktyp^{\phantom{ty}2s}$.
}}
 Therefore $\Gamma_3 \propto b^{2s}$ is small in the zero-range regime, as anticipated.

\item
  {\bl {\bl Based on heuristic arguments,} it was already {\bl stated} 
    in~\cite{Petrov4body2004} (see also~\cite{PSS_PRA,PetrovJPhysB,PetrovLesHouches2010}) that in the zero-range regime, the formation rate of deeply bound dimers is
    proportional to the probability
    of finding three particles at distances $\lesssim b$, times $\hbar / (m\,b^2)$, that is,
    $\Gamma_3 = \mathcal{K}\,N_3(b)\,\hbar/(m\,b^2)$, with $N_3(b)$ evaluated within the zero-range model, and  $\mathcal{K}$ a dimensionless prefactor that depends on short-range three-body physics. This statement is equivalent to (\ref{eq:Gamma3}), given the relation~(\ref{eq:N3}), with $\mathcal{K} = - 8\,s\,(s+1)\,{\rm Im}\,a_3 / b^{2s}$.
    The novelties of the present work are
    {\it (i)}
    to provide a derivation of the relation (\ref{eq:Gamma3}), and hence of the above statement from~\cite{Petrov4body2004},
    {\it (ii)}
    to introduce the natural single parameter $a_3$ (instead of the two parameters $\mathcal{K}$ and $b$), and
    {\it (iii)}
    to generalize the relation to more complex interactions (in the following Section~\ref{subsec:gamma3_real}).   
}

\item 
Let us consider the case of the homogeneous unpolarized {\bl zero-temperature} 
unitary gas,
of number density $n$. 
Introducing the three-body contact density $\Cr_3 := C_3 / \Vr$ with $\Vr$ the volume,
dimensional analysis gives
$\Cr_3 \ =\  \zeta_3\ n^{(2s+5)/3}$
where $\zeta_3$ is a dimensionless constant.
Hence
\be
-\dot{n} = 24\,s\,(s+1)\,\frac{\hbar}{m}\ (-{\rm Im}\,a_3)\ \zeta_3\ n^{(2s+5)/3}\,.
\label{eq:ndot}
\ee
Therefore, as already found in~\cite{Petrov4body2004},
the timescale of three-body losses $\tau_3 := n/ |\dot{n}|$
is of order $\tau_F / (k_F b)^{2s}$,
much larger than the thermalization time $\tau_F \sim m / (\hbar\,k_F^2)$,
{\bl so that the gas remains at quasi-equilibrium.
  }

\item {\brown For bosons (and more generally in presence of the Efimov effect) it is commonly accepted that three-body losses can be described by making the three-body parameter complex~\cite{RevueBraaten2,Braaten_etats_d_efimov,RevueBraaten}.
  We have transposed this to the fermionic case (where the Efimov effect does not occur) by introducing a complex three-body parameter $a_3$. The expression~(\ref{eq:Gamma3}) of \,$\Gamma_3$ is reminiscent of the relation for bosons expressing \,$\Gamma_3$ in terms of $C_3$ and the inelasticity parameter ({\it i.e.} the phase of the three-body parameter)~\cite{WernerCastinRelationsBosons,BraatenLindblad}.
  An important difference is that in the fermionic case, in the zero-range regime, the three-body parameter is small, and can be set to zero when evaluating typical observables 
 other than~$\Gamma_3$.}
\item {\bl The notion of three-body parameter $a_3$ {\brown differs from} the three-body scattering hypervolume $D$ which was defined for bosons in 3D~\cite{TanD,ZhuTanD} and for various other cases~\cite{TanD_uneq_masses,TanDfermions,TanDfermions2D,TanDfermions1D}. Presumably, $D$ could be defined also for the present case (two-component fermions in 3D),
  and as in~\cite{TanD,ZhuTanD,TanD_uneq_masses,TanDfermions,TanDfermions2D,TanDfermions1D}, $D$~would govern the asymptotic behavior of the three-body zero-energy wavefunction at distances $ \gg |a_2|$
 (while $a_3$ governs the regime of distances $\ll |a_2|$ and $\gg b$)
and $\Gamma_3$ would have a simple expression in terms of $\,{\rm Im}\, D$  in the weakly interacting regime $\ktyp |a_2| \ll 1$
(whereas (\ref{eq:Gamma3}) 
remains valid in the strongly correlated regime $\ktyp |a_2| \gtrsim 1$).
{\brown On the other hand, $a_3$ is only defined for resonant interactions ($|a_2| \gg b$) while $D$ also exists for non-resonant interactions.}}

\ei

We turn to the derivation of (\ref{eq:Gamma3},\ref{eq:Gamma2,1},\ref{eq:Gamma1,2}).
{\bl Our reasoning is similar to the bosonic case treated in App.~B of~\cite{WernerCastinRelationsBosons}, but the present case is significantly more complicated 
because we cannot work directly within the zero-range model.}
For $R\gg b$, the 
$a_3/R^s$ term in (\ref{eq:def_a3}) is negligible compared to the
$R^s$ term,
consistently with the behavior (\ref{eq:R0}) within the zero-range model.
However this 
$a_3/R^s$ term cannot be neglected to describe three-body losses,
since the losses come from 
the non-zero imaginary part of $a_3$, as we will see.
{\bl Accordingly, for the Gamow state $\psi$, we} go beyond the zero-range model
and replace (\ref{eq:R0}) with
\be
\psi(\rr_1,\ldots,\rr_N)
\simeq
\left(R^{s}
- \frac{a_3}{R^s}
\right)
\,\frac{1}{R^2}\
\sum_{\md=-1}^{+1}\ \phi_\md(\Oo)\ B_\md(\CC;\rr_4,\ldots,\rr_N)
\label{eq:R0_b}
\ee
which we expect to hold provided
\bea
R & \leq & \ddd
\label{eq:R<d}
\\
b &\ll& r_{ij} 
,\ \forall i<j {\bl \leq 3}
\label{eq:b<<rij}
\\
  {\bl \ddd}
  & {\bl \ll } & {\bl r_{ij} ,\ \forall i<j,\ j\geq4}
  \label{eq:d<<rij}
\eea
where $\ddd$ is a length that satisfies
\bea
b &\ll& \ddd 
\label{eq:<<d}
\\
\ddd &\ll& {\bl {\rm Min}\left(|a_2|, 1/\ktyp\right)}
\label{eq:d<<}
\eea
and whose appropriate choice will be discussed more precisely below.
{\bl
The purpose of (\ref{eq:d<<rij}) is to ensure that the interparticle distances within the triplet of particles $1,2,3$ are much smaller than all other interparticle distances.}




{\bl We use} the shorthand notation
$\XX = (\rr_1,\ldots,\rr_N)$.
{\bl The solution $\Psi(\XX;t)$ of the time-dependent Schr\"odinger equation [$i\hbar \,\dot{\Psi} = H\,\Psi$, with $H$ given by (\ref{eq:H_V2})] associated with the Gamow state $\psi(\XX)$ is 
  \be
  \Psi(\XX; t) = \psi(\XX)\,e^{- i E t / \hbar}\,.
  \label{eq:Psi_t}
  \ee
It satisfies the continuity equation
\be
\partial_t\left(|\Psi|^2\right) = - \gr_\XX \cdot \JJ
\label{eq:continu}
\ee
in terms of the probability current
\be
\JJ := \frac{\hbar}{m}\ {\rm Im}\, \left(\Psi^* \gr_\XX \Psi \right)\,.
\label{eq:current}
\ee}
{\bl We express the loss event rate
  ({\it i.e.} the probability for a 
  loss event to occur per unit of time) as
\be
\Gamma = - \left.\frac{\partial}{\partial t}\right|_{\bl t=0}\ \int_\Rr d^{3N}\!X \ \, |\Psi(\XX; t)|^2
\label{eq:Gamma3_Norm}
\ee
where the region $\Rr$ should physically correspond to $N$ atoms in the trap, 
and where $\psi(\XX) = \Psi(\XX,t=0)$ is normalized by the condition
\be
\int_\Rr d^{3N}\!X \ \, |\psi(\XX)|^2 = 1.
\label{eq:norm}
\ee

There is some freedom in how to define $\Rr$.
Let $\Rr_{\rm trap}$ denote the ``trapping region'' of the potential $U(\rr)$, which can be defined as the set of points $\rr$ such that the classical trajectory of a particle with initial position $\rr$ and zero initial velocity remains bounded.\footnote{\bl For example, if $U(\rr) = m\,\omega^2 r^2 - \lambda\,r^4$, then $\Rr_{\rm trap}$ is the sphere centered at the origin of radius $r_{\rm trap}$ such that $U(r)$ reaches its maximum at $r=r_{\rm trap}$, {\it i.e.} $r_{\rm trap} = \sqrt{m/\lambda}\times \omega/2$. This definition of $\Rr_{\rm trap}$ is merely one particularly simple and natural choice among a range of possibilities.}
A possible definition of $\Rr$ would be $\Rr = (\Rr_{\rm trap})^N$,
but for later convenience, we choose $\Rr$ to be slightly smaller, 
by excluding configurations with two or three nearby particles:
\be
\Rr = (\Rr_{\rm trap})^N \ \setminus \  (\Rr_2 \cup \Rr_3)
\nonumber
\ee
where 
  \bea
  \Rr_2 &=& \bigcup_{i:\UP, j:\down} B_{ij}  \ \ \ \ \ {\rm with}\ \ \ B_{ij} \,=\, \left\{ \, (\rr_1 \, , \, \ldots \, , \, \rr_N) \, \big| \ r_{ij} < \dd \, \right\}
  \\
  \Rr_3 &=& \bigcup_{i<j:\UP, k:\down \ {\rm or} \ i<j:\down, k:\UP} B_{ijk}
  \ \ \ \ \ {\rm with}\ \ \ B_{ijk} \,=\, \left\{ \, (\rr_1 \, , \, \ldots \, , \, \rr_N) \, \big| \ R_{ijk} < \ddd \, \right\}\,.
\label{eq:defR3}
  \eea
In other words, $(\rr_1 , \ldots , \rr_N) \in \Rr$ means that
\bi
\item
  all particle positions $\rr_i$ are inside the trapping region $\Rr_{\rm trap}$ 
\item
  all distances between pairs of particles with opposite spin are $\geq\dd$ 
\item
  all hyperradii of triplets of particles with non-identical spins are $\geq\ddd$.
  \ei
  We take $\dd$ such that
  \bea
  b &\ll& \dd
  \\
  \dd &\ll& \ddd\,. \label{eq:d2<<d3}
  \eea
  Conditions (\ref{eq:d<<},\ref{eq:d2<<d3}) ensure that both $\dd$ and $\ddd$ are $\ll 1/\ktyp$. Hence, to leading order, 
  the normalization integral (\ref{eq:norm}) is independent of $d_2$ and $d_3$,
  because $\Rr_2$ and $\Rr_3$ are negligibly small subsets of $(\Rr_{\rm trap})^N$. Furthermore, to leading order in the zero-range regime,
  \bi
\item there exists a stationary state of the zero-range model
whose wavefunction (normalized by the usual integral over the entire space) is close to the Gamow state $\psi(\XX)$ for $\XX \in \Rr$
\item
  {\bl in (\ref{eq:R0_b}), $B_m(\CC; \rr_4 , \ldots \rr_N)$ can be evaluated within the zero-range model, provided $\CC, \rr_4 , \ldots \rr_N$ all belong to $\Rr_{\rm trap}$;
  indeed, the $\propto a_3$ term is a small correction, and the condition (\ref{eq:R0_b}) for the Gamow state must match with the condition (\ref{eq:R0}) for the zero-range model.}
    \ei

Equations~(\ref{eq:Psi_t},\ref{eq:Gamma3_Norm},\ref{eq:norm}) directly yield
\be
\Gamma = -\frac{2}{\hbar}\ {\rm Im}\, E\,,
\label{eq:ImE}
\ee
which we will use for a consistency check in Sec.~\ref{sec:dE}.
Here we will evaluate 
the loss rate by a flux computation.
We use the notation
\be
\Phi(\Sr) := \int_{\Sr}{\bf d^{3N-1}  S} \ \cdot \ {\bf J}
\label{eq:defPhi}
\ee
for the probability flux through a surface $\Sr$.
In~(\ref{eq:Gamma3_Norm}), we interchange the time derivative and the integration, and use the continuity equation (\ref{eq:continu}) and
the divergence theorem, which yields
\be
\Gamma = \Phi(\partial \Rr)
\label{eq:Gamma3_J}
\ee
where $\partial \Rr$ is the boundary of $\Rr$. Here and in what follows, the differential surface vector ${\bf d^{3N-1}S}$ appearing in (\ref{eq:defPhi}) is oriented towards the exterior of $\Rr$.
Assuming that the trap depth is large enough for evaporation to be negligible\footnote{\bl \label{fn:evap} 
\bl Evaporation is the process of an individual atom (or a weakly bound dimer for small positive $a_2$) escaping directly 
from the trapping region (because its energy is large due to a rare fluctuation, and/or it tunnels through the trapping-potential barrier). This process is exponentially suppressed
in the limit where the trap depth is large compared to the typical energy per atom (subtracting the dimer binding energy contribution for $a_2>0$). 
For evaporation of individual atoms, the trap depth can be defined as $[{\rm Min}_{\rr\,\in\,\partial\Rr_{\rm trap}} U(\rr)]-[{\rm Min}_{\rr\,\in\,\Rr_{\rm trap}} U(\rr)]$,
the evaporation rate is $\Phi(\Sr_{\rm trap})$ with
$\Sr_{\rm trap} = \partial\big( (\Rr_{\rm trap})^N\big) \cap \Rr = \big\{ \XX\in \Rr \, \big| \, \exists i,\ \rr_i\in \partial\Rr_{\rm trap} \big\}$,
and this rate is exponentially suppressed because 
$\JJ(\XX)$ 
with $\XX \in \Sr_{\rm trap}$ is exponentially suppressed.},
we have
\be
\Gamma = 
\Phi(\Sr_2) + \Phi(\Sr_3)
\ee
where
\be
\Sr_2 = \partial\Rr_2 \cap \Rr\,,
\ \ \ \ \ \ \ \ \ \ \ \ 
\Sr_3 = \partial\Rr_3 \cap \Rr\,.
\nonumber
\ee
{\gre A visual representation is shown in Figure~\ref{fig:illust}.}
On physical grounds, we identify
$\Phi(\Sr_2)$ as the two-body loss rate $\Gamma_2$,
and $\Phi(\Sr_3)$ as the three-body loss rate $\Gamma_3$\,,
which determine the average number of\, lost atoms per unit of time: $-\dot{N} = 
2\,\Gamma_2 + 3\,\Gamma_3$.
In the considered regime of small range and large trap depth, we expect that $\Gamma_2$ is given by the relation (\ref{eq:Gamma2}),
and that in the considered case where $a_2$ is real, $\Gamma_2$ is negligible compared to $\Gamma_3$
so that
$\Gamma \simeq \Gamma_3$\,.\footnote{\bl Indeed, in the zero-range limit, the reasoning of footnote~\ref{fn:Gamma2} \ yields the expression (\ref{eq:Gamma2}) for $\Phi(\Sr_2)=\Gamma_2$\,; moreover, in the case where $a_2 \in \mathbb{R}$, we expect $\Gamma_2 \ll \Gamma_3$\,, because there is no mechanism that would generate a non-negligible flux
exiting $\Rr$ through $\Sr_2$
(hence entering into $\Rr_2$) and propagating in $\Rr_2$ with an initial wavevector high enough to climb the trapping potential barrier and escape from $(\Rr_{\rm trap})^N$.}

\begin{figure}
  \includegraphics[width=0.9\columnwidth]{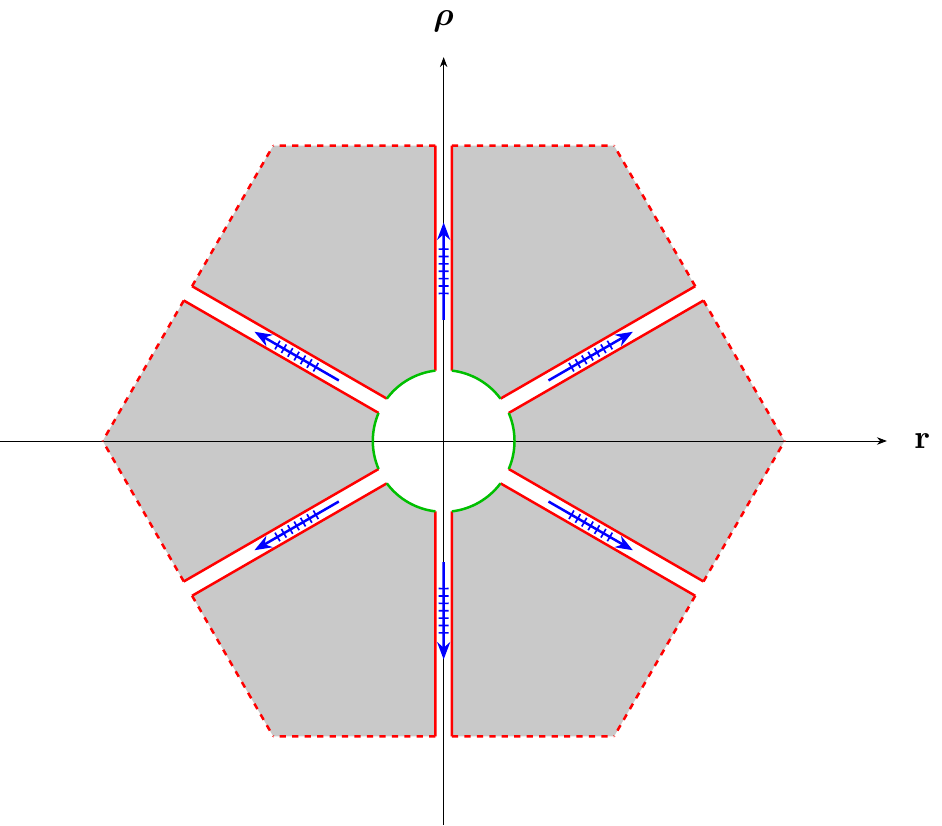}
  \caption{
   Geometric illustration of the three-body loss process.
    The total decay rate is given by the probability flux exiting from the region~$\Rr$ (grey shaded area).
Neglecting the flux through $\Sr_2$ (red straight lines) which corresponds to two-body losses, and the flux through $\Sr_{\rm trap}$ (dashed red lines) which corresponds to evaporation,
the dominant contribution is the flux through $\Sr_3$ (green circular arcs) which corresponds to three-body losses.
    The blue arrows represent the {\it deep-dimer + atom} outgoing wave, corresponding to a deeply bound dimer and an atom flying apart with a large relative momentum and escaping from the trap (this wave propagates in the region~$\Rr_2$).
    For the purpose of making this illustrative drawing two-dimensional, we considered $N=3$ particles in one space dimension, and fixed the center-of-mass coordinate to $\CC=\vn$. The positions of the three particles are then determined by the Jacobi coordinates $\rr$ and $\rrho$. The trapping region was simply assumed to be a symmetric interval around the origin.
    \label{fig:illust}}
\end{figure}

It remains to determine $\Gamma_3  = \Phi(\Sr_3)$. We will use the notation
\be
\Sr^{(ijk)} \ := \ \partial B_{ijk} \, \cap \, \Rr\, = \big\{ \, \XX \in \Rr \, \big| \,
R_{ijk}
= d_3 \big\}\,.
\nonumber
\ee
From (\ref{eq:defR3}), we have
$\ds \Sr_3 = \bigcup_{i<j:\UP, k:\down \ {\rm or} \ i<j:\down, k:\UP} \Sr^{(ijk)}$.
Hence $\Phi(\Sr_3) = \Gamma_{2,1} + \Gamma_{1,2}$ with
\bi
\item $\ds \Gamma_{2,1} = \sum_{i<j:\UP, k:\down} \Phi\big(\Sr^{(ijk)}\big)$ the contribution from $\UP\UP\down$ triplets 
\item $\ds \Gamma_{1,2} = \sum_{i<j:\down, k:\UP} \Phi\big(\Sr^{(ijk)}\big)$ the contribution from $\UP\down\down$ triplets.
  \ei}
\noindent By antisymmetry, each $\UP\UP\down$ triplet gives the same contribution, so that
\be
\Gamma_{2,1} 
= 
\frac{N_\UP \, (N_\UP - 1)\,N_\down}{2}\,
\ {\bl \Phi\big(\Sr^{(123)}\big)}
\,.
\label{eq:Gamma3_J2}
\ee
This can be rewritten as \footnote{
{\bl To justify this rewriting, we mostly follow the bosonic case treated in App.~B of~\cite{WernerCastinRelationsBosons}} 
  {\bl (we will also correct in passing a minor error in an intermediate step in~\cite{WernerCastinRelationsBosons}).}
We need to evaluate $Q := \int_{\Sr^{(123)}} \left(
\psi^* \, \gr_\XX \psi
-
\psi \, \gr_\XX \psi^*
\right) \cdot {\bf d^{3N-1}S}$\,.
The constraint $R \equiv R_{123} > \ddd$ which defines
the domain $B_{123}$ does not impose any constraint on $(\rr_4,\ldots,\rr_N)$. Therefore the differential surface vector ${\bf d^{3N-1}S}$, being normal to $\partial B_{123}$, only has its 9 first coordinates which are non-zero, while its coordinates 10 to $3N$ are vanishing.
More precisely, {\bl denoting by ${\bf d^{3N-1}S}_t$ the 9-dimensional vector whose coordinates are equal to the 9 first coordinates of ${\bf d^{3N-1}S}$, we have ${\bf d^{3N-1}S}_t = {\bf d^{8}S}\ \dr_4 \ldots \dr_N$}
where ${\bf d^{8}S}$ is the differential surface vector of the domain 
$\{(\rr_1,\rr_2,\rr_3) \ / \ R>\ddd \}$.
Hence 
$Q = \int_{\Sr^{(123)}} \,\dr_4 \ldots \dr_N\ 
{\bf d^{8}S} \cdot \left(
\psi^* \, \gr_{\tilde{\XX}} \psi
-
\psi \, \gr_{\tilde{\XX}} \psi^*
\right)$
where ${\tilde{\XX}} := (\rr_1,\rr_2,\rr_3)$.
Applying the divergence theorem backwards, this can be rewritten
$Q = \int_{\Sr^{(123)}} \,\dr_4 \ldots \dr_N\ 
d^9\tilde{X}\ \ \gr_{\tilde{\XX}} \cdot \left(
\psi^* \, \gr_{\tilde{\XX}} \psi
-
\psi \, \gr_{\tilde{\XX}} \psi^*
\right)$.
We then perform the change of variables $\ \tilde{\XX} \longrightarrow (\CC,\RR)\ $,
and rewrite the integrand as
${\bl \psi^* \Delta_{\tilde{\XX}} \psi - \psi \, \Delta_{\tilde{\XX}} \psi^* =
\frac{1}{3} \left( \psi^* \Delta_{\CC} \psi - \psi \, \Delta_{\CC} \psi^* \right)
+ 2 \left( \psi^* \Delta_{\RR} \psi - \psi \, \Delta_{\RR} \psi^* \right)
=}
\frac{1}{3}\ \gr_\CC \cdot 
\left(
\psi^* \, \gr_{{\CC}} \psi
-
\psi \, \gr_{{\CC}} \psi^*
\right)
+ {\bl 2} \ \gr_\RR \cdot \left(
\psi^* \, \gr_{{\RR}} \psi
-
\psi \, \gr_{{\RR}} \psi^*
\right)$.
By the divergence theorem,
the integral over $\CC$ of the $\gr_\CC$ term vanishes,
while the integral over $\RR$ of the $\gr_\RR$ term yields the result
$Q = -\frac{3\sqrt{3}}{4}\ \ddd^5 \ \int_{\Sr^{(123)}} \dC\ \dr_4 \, \ldots \, \dr_N
\ d^5\Omega\
\left(
\psi^* \, \partial_R 
\psi
-
\psi \, \partial_R 
\psi^*
\right)$\,.}
\be
\Gamma_{2,1} = -
N_\UP \, (N_\UP - 1)\,N_\down\,
\frac{3\sqrt{3}}{8}\ \ddd^5 
\int_{\bl \Sr^{(123)}}\  \dC\  \dr_4 \, \ldots \, \dr_N \ d^5\Omega
\ \ \frac{\hbar}{m} \  {\rm Im} \left[ \psi^* \partial_R \psi \right]\,. 
\label{eq:Gamma3_partial}
\ee
{\bl We then simplify this expression in two steps:}
  \bi
     \item[\bl {\it (i)}] replace $\psi$ with its asymptotic behavior
(\ref{eq:R0_b}), with $B_\md$ evaluated within the zero-range model
{\bl \item[{\it (ii)}] replace the integration domain $\Sr^{(123)}$ by the 
  entire region $\partial B_{123}$.}
\ei
   {\bl
Step {\it (i)} is justified since 
the conditions (\ref{eq:R<d},\ref{eq:b<<rij},\ref{eq:d<<rij}) hold except in a negligibly small domain of the integration variables $\CC, \rr_4, \ldots , \rr_N, \Oo$.
Step {\it (ii)} is justified because given (\ref{eq:d2<<d3}), the condition $\XX\notin \Rr_2$ only excludes a negligibly small region of hyperangles $\Oo$.
We note that the order of these two steps is important.\footnote{\bl If we would start by
replacing $\Sr^{(123)}$ with $\partial B_{123} \, \cap \, (\Rr_{\rm trap})^N$ (keeping the original Gamow-state wavefunction~$\psi$),
then we would get a vanishing result for the total flux : The flux through 
$\partial B_{123} \, \cap \, (\Rr_{\rm trap})^N \setminus \Rr_2$
(corresponding to the three-atom wave partially reflected from the small-$R$ region)
would be compensated by the flux through the complementary surface $\Rr_2 \, \cap \, \partial B_{123} \, \cap \, (\Rr_{\rm trap})^N$ where the {\it deep-dimer + atom}\, outgoing-wave contribution to $\psi$ gives the main contribution to the flux. This follows from the fact that for the three-body zero-energy scattering state $\Psi_{\md}$, the total flux $\int d^5 \Omega \ \, {\rm Im}(\Psi_{\md}^*\,\partial_R \Psi_{\md})$ vanishes (a related discussion can be found in~\cite{ZhuTanD}).}}
Using (\ref{eq:C3_BB}) then yields the result (\ref{eq:Gamma2,1}).
The expression (\ref{eq:Gamma1,2}) of $\Gamma_{1,2}$ is derived analogously.

Let us now discuss in more detail
the appropriate choice of the length $\ddd$. 
The condition~(\ref{eq:<<d}) 
is actually not sufficient in order to have the behavior~(\ref{eq:R0_b}) of $\psi$.
Indeed, we expect
[based on the small-$R$ expansion of the finite-energy solution $J_s(kR)$ of the hyperradial Schr\"odinger equation~(\ref{eq:schro_R})]
that in addition to the term $R^s$ in (\ref{eq:R0_b}),
there is a higher-order correction term of order $R^{s+2}\ktyp^2$,
which is negligible compared to $a_3/R^s$ provided we take
\be
\ddd \ {\bl \ll}\  b / (\ktyp b)^{1/(s+1)}.
\label{eq:l_2}
\ee
The condition {\bl $\ddd \ll 1/\ktyp$} is then automatically satisfied.
Compatibility with (\ref{eq:<<d}) then requires $(\ktyp b)^{1/(s+1)} \ll 1$,
which is a quite stringent condition given the smallness of the exponent $1/(s+1)\simeq 0.36$.
However, while this condition is necessary for~(\ref{eq:R0_b}),
{\bl we do not expect it to} be necessary for the final expression~(\ref{eq:Gamma3}) of the three-body loss rate.

{\bl To complete our discussion of validity conditions, we now consider the contribution of the angular-momentum sector $l=0$.} Another condition to fulfill in order for~(\ref{eq:R0_b}) to be valid is that one can neglect the contribution
coming from the $l=0$ sector 
of the unitary three-body problem.
Let us denote by $s':= s_{l=0, n=0} = 2.166221977\ldots$ the smallest solution different from~2 of $s' \cos(s'\pi/2) + 4\,\sin(s'\pi/6) /\sqrt{3}=0$,
and by $\phi'(\Oo) := \phi_{(l=0,n=0)}(\Oo)$ the corresponding $l{=}0$ hyperangular wavefunction.
There is a higher-order correction to the r.h.s. of~(\ref{eq:R0_b}) given by
$R^{s'-2}\,\phi'(\Oo)$ times a function $B'(\CC;\rr_4,\ldots,\rr_N)$.
Requiring this ${\propto}R^{s'}$ correction to be negligible compared to the ${\propto}a_3/R^s$ term in (\ref{eq:R0_b}) would yield an additional condition on $\ddd$,
but this 
is not necessary for the final result~(\ref{eq:Gamma3}).
Instead, what is truly necessary for~(\ref{eq:Gamma3})
is that the contribution from the $l{=}0$ angular-momentum sector to $\Gamma_3$ is negligible compared to the $l{=}1$ contribution.
Adding the term
$(R^{s'} - a'_3/R^{s'})\,R^{-2}\,\phi'(\Oo)\,B'(\CC;\rr_4,\ldots,\rr_N)$ to the r.h.s. of~(\ref{eq:R0_b}),
and still using~(\ref{eq:Gamma3_partial}),
we obtain the additional term 
$\Gamma'_3 = -(\hbar/m)\,{\bl 4} s' (s'+1) \, C'_3\ {\rm Im} \, a'_3$
on the r.h.s. of~(\ref{eq:Gamma3}).
Here $C'_3$ is the ``$l{=}0$ three-body contact'',
defined in such a way that there is a higher-order correction $C'_3\, \epsilon^{2 s' + 2}$
to the r.h.s. of~(\ref{eq:N3}).
The ratio $\Gamma'_3 / \Gamma_3 \propto b^{2(s'-s)}$ is small in the zero-range regime (except if $C_3$ is anomalously small).
For example, for the degenerate unpolarized unitary gas,
this yields the truly necessary condition $(n^{1/3} b)^{2(s'-s)}\ll1$ for the validity of~(\ref{eq:Gamma3}).


\subsubsection{{\bl General} interactions}
\label{subsec:gamma3_real}

{\bl In cold atom experiments, interactions are more complex than the simple model of Sec.~\ref{subsec:simple}.
Not only two-body interactions, but also three-body interaction are present.
Moreover,
\begin{itemize}
  \item interactions are not necessarily rotationally invariant around any axis, due to the presence of the external magnetic field
  \item interactions are not necessarily symmetric w.r.t. exchanging the role of $\UP$ and $\down$.
    \end{itemize}
As a minimal model including these features, we consider a two-body interaction potential $V_2(\rr)$ which may now depend on the direction of $\rr$,}
and a three-body interaction potential $V_{2,1}$ (resp.~$V_{1,2}$) between triplets of particles of spins $\UP\UP\down$ (resp. $\UP\down\down$).
\\The corresponding stationary $N$-body Schr\"odinger equation is
\be
\sum_{i=1}^N \left[-\frac{\hbar^2}{2m}\Delta_{\rr_i} + U(\rr_i) \right] \psi
\ +\  \sum_{\substack{i:\UP,\,j:\down}} V_2(\rr_{ij}) \ \psi
\ +\  \sum_{\substack{i<j:\UP,\,k:\down}} V_{2,1}(\RR_{ijk}) \ \psi
\ +\  \sum_{\substack{i<j:\down,\,k:\UP}} V_{1,2}(\RR_{ijk}) \ \psi
 = E\,\psi\,.
\label{eq:schro_V3}
\ee
Here $\RR_{ijk}$ denotes the Jacobi coordinates associated to particles $i,j,k$ [defined by replacing the indices 1,2,3 by $i,j,k$ in (\ref{eq:def_jaco},\ref{eq:def_RR})].
\\The two-body interaction $V_2$ is still assumed to be resonant, 
$|a_2| \gg b_2$ (with $a_2$ and $b_2$ the scattering length and the range of $V_2$).
The three-body interaction potential is assumed to have a finite range~$b_3$,
in the sense that it decays quickly at hyperradii larger than $b_3$.
\\We consider the zero-range regime where
\be
b := {\rm Max}(b_2, b_3)
\nonumber
\ee
is much smaller than $|a_2|$ and $1/\ktyp$,
where $1/\ktyp$ is defined as the smallest scale of variation of the wavefunction $\psi$ in the region where all interparticle distances are $\gg b$. For alkali atoms near an open-channel dominated Feshbach resonance,
$b_2$ is set by the van~der~Waals length~\cite{FeshbachRMP2010},
which is $\ll 1/\ktyp$ in typical cold-atom experiments.

Instead of a single three-body parameter $a_3$, there are
in general six three-body parameters $a^{(\md)}_{2,1}$ and~$a^{(\md)}_{1,2}$,
with $\md \in \{ -1,\ 0,\ 1\}$ the angular momentum quantum number.
\\We find that the $\UP\UP\down$ and $\UP\down\down$ three-body loss-rates are given by
\be
\boxed{\Gamma_{2,1}  \simeq 
-\frac{\hbar}{m}\ 8 s (s+1) \, \sum_{\md=-1}^1 C^{(\md)}_{2,1} \ {\rm Im} \,
a^{(\md)}_{2,1}}
\label{eq:Gamma2,1_real}
\ee
\be
\boxed{\Gamma_{1,2}  \simeq 
-\frac{\hbar}{m}\ 8 s (s+1) \, \sum_{\md=-1}^1 C^{(\md)}_{1,2} \ {\rm Im} \,
a^{(\md)}_{1,2}.}
\label{eq:Gamma1,2_real}
\ee
Here, the $\md$-resolved three-body contacts $C^{(\md)}_{2,1}$ and $C^{(\md)}_{1,2}$ are defined by
\be
C^{(\md)}_{2,1} := 
N_\UP (N_\UP-1) N_\down
\ \frac{3\sqrt{3}}{32\,(s+1)}\  
\int |B_\md(\CC;\rr_4,\ldots,\rr_N)|^2
d^3\!C\, \dr_4 \ldots \dr_N
\label{eq:C3m_BB}
\ee
where $B_\md$ is related 
to the many-body wavefunction (in the zero-range limit) through Eq.~(\ref{eq:R0}),
and similarly
\be
C^{(\md)}_{1,2} := 
N_\down (N_\down-1) N_\UP
\ \frac{3\sqrt{3}}{32\,(s+1)}\ 
\int |\tilde{B}_\md(\CC;\rr_2,\rr_5,\ldots,\rr_N)|^2
d^3\!C\, \dr_2 \, \dr_5 \ldots \dr_N
\label{eq:C3m_BBtilde}
\ee
where $\tilde{B}_\md$ is defined in Eq.~(\ref{eq:R0t}).
\\Note that from Eqs.~(\ref{eq:C3_BB},\ref{eq:C3_BBt}) we have
$\sum_{\md=-1}^1 C^{(\md)}_{2,1} = C_{2,1}$
and $\sum_{\md=-1}^1 C^{(\md)}_{1,2} = C_{1,2}$.

The three-body parameters $a_{2,1}^{(\md)}$ are defined as follows.
Setting
\be
\Vr(\RR) \ :=  \ V_2(\rr_{13}) + V_2(\rr_{23}) + V_{2,1}(\RR),
\label{eq:def_Vr}
\ee
the solution $\Psi_\md$
of the zero-energy Schr\"odinger equation in free space for two $\UP$ and one $\down$ particles
\be
-\frac{\hbar^2}{m} \ \Delta_{\RR} \Psi_\md \ +\  \Vr(\RR)\ \Psi_\md \ =\ 0
\label{eq:schro3_V3}
\ee
with angular-momentum quantum numbers $(l=1,\md)$
has the asymptotic behavior 
\be
\boxed{\Psi_\md(\RR) 
\simeq
\left(
R^s - \frac{a_{2,1}^{(\md)}}{R^s}
\right)
\frac{1}{R^2}\ 
\phi_\md(\Oo)}
\label{eq:a21}
\ee
in the region where all interparticle distances are $\gg b$ and $\ll |a_2|$.
The three-body parameters $a_{1,2}^{(\md)}$ are defined similarly in terms of the $\UP\down\down$ three-body scattering states.

The relation~(\ref{eq:Gamma2,1_real})
 is derived by considering the probability current, in a completely analogous way to Eqs.~(\ref{eq:Gamma3_Norm},\ref{eq:Gamma3_J},\ref{eq:Gamma3_J2},\ref{eq:Gamma3_partial}) above, using
 the asymptotic behavior of the many-body wavefunction
 in the region (\ref{eq:R<d},\ref{eq:b<<rij},\ref{eq:d<<rij}) which is now given by
\be
\psi(\rr_1,\ldots,\rr_N)
\simeq
\frac{1}{R^2}
\ \sum_{\md=-1}^{+1}
\left(R^{s}
- \frac{a_{2,1}^{(\md)}}{R^s}
\right)
\ \phi_\md(\Oo)\ B_\md(\CC;\rr_4,\ldots,\rr_N).
\label{eq:R0_real}
\ee
The expression (\ref{eq:Gamma1,2_real}) of $\Gamma_{1,2}$ is derived analogously.

\vskip 0.17cm
\noindent \underline{\it Discussion:} 
\\ The parameters
${\rm Im}\,a^{(\md)}_{2,1}$ and  ${\rm Im}\,a^{(\md)}_{1,2}$
are {\it a priori} unknown.\footnote{One may expect a small relative difference between $a^{(0)}_{i,j}$ and $a^{(\pm 1)}_{i,j}$,
  and an even smaller one between $a^{(1)}_{i,j}$ and~$a^{(-1)}_{i,j}$,
  similarly to the $\md$-dependence of the two-body $p$-wave scattering volume not too close to a $p$-wave Feshbach resonance~\cite{BohnPwaveMultiplet,
    WeidemullerPwaveTripletShort,WeidemullerPwaveTripletLong
  }.} 
In principle, one may hope to compute them theoretically by solving
a sufficiently realistic three-body problem, but this is a difficult task~\cite{WangJulienneMultichan3body,Cornell_3BP_vdw,KokkelmansMultichan3body}.\footnote{For such a computation of the three-body parameters, one may need to take into account that an atom has more than two relevant internal states $|\nu\ra$. However, we expect the general relations (\ref{eq:Gamma2,1_real},\ref{eq:Gamma1,2_real}) to remain applicable. Indeed, we expect that in typical experiments,
the atoms mainly occupy two internal states, that we can label $\nu = \, \UP$ and $\down$, and if all interatomic distances $r_{ij} \gg b$,
the many-body wavefunction
$\Phi(\rr_1,\nu_1; \ldots ; \rr_N, \nu_N)$ is non-negligible only
if all $\nu_i$ belong to $\{ \UP ; \down \}$,
in which case $\Phi$ is given to good accuracy by
antisymmetrizing the wavefunction $\psi(\rr_1, \ldots , \rr_N)$ of the zero-range model.}
Instead, one could determine them by measuring 
the three-body loss rate in situations where the three-body contacts
$C^{(\md)}_{2,1}$ and  $C^{(\md)}_{1,2}$ are
known theoretically.

A~first possibility is to work
with a small number $N$ of particles in a microtrap~\cite{Andersen3atomLoss,JochimFewFermion,JochimFewToMany,JochimPairingFew} where the $N$-body wavefunction can be computed numerically with good accuracy~\cite{BlumeCRAS,BlumeRelations4fermions,BlumeRevue} so that the three-body contacts $C_{i,j}^{(\md)}$ could be calculated.
Measuring $\Gamma_3$ in six different states and inverting (\ref{eq:Gamma2,1_real},\ref{eq:Gamma1,2_real}) would allow to determine the six parameters
${\rm Im}\,a^{(\md)}_{i,j}$.
The case of three particles in an isotropic harmonic trap is particularly simple: The problem is analytically solvable~\cite{TanScalingREVTEX,Werner3corpsPRL}, and
if one prepares one of the
six degenerate ground states
corresponding to a given quantum number $\md \in \{ -1; 0;  1 \}$
and $(N_\UP,N_\down) \in \{ (2,1) ; (1,2) \}$,
then $C^{(\md)}_{N_\UP,N_\down}$ is the only non-zero three-body contact,
so that $\Gamma_3$ is simply proportional to ${\rm Im}\,a^{(\md)}_{N_\UP,N_\down}$.
{\bl Explicitly, taking for example $N_\UP=2$ and $N_\down = 1$, we get
$C_3=C^{(\md)}_{2,1} = (m\omega/\hbar)^{s+1}  \, / \, [2^{s+1}\,\Gamma(s+2)]\,$,
hence
$\Gamma_3 / \omega = -{\rm Im}\,a^{(\md)}_{2,1}\,(m\omega/\hbar)^s \times
1.266322$
(in agreement with the scaling given in~\cite{Werner3corpsPRL}).\footnote{\bl For the excited state whose energy is $2q\,\hbar\omega$ above the ground state energy, the value of $C^{(\md)}_{2,1}$ is multiplied by ${q+s \choose s} = \frac{\Gamma(q+s+1)}{\Gamma(s+1)\,q!}$ (which is a growing function of $q$, meaning that the growing
delocalization
in the trap is overcompensated by the growing penetration under the $s^2/R^2$ barrier).}}


A second possibility is to work
with a homogeneous (or locally homogeneous) unpolarized gas at
equilibrium,
for which the six three-body contacts $C^{(\md)}_{i,j}$ are 
all equal, as shown in Appendix~\ref{app:C3m}.
This gives
\be
\boxed{\Gamma_3 \simeq -\,\frac{\hbar}{m}\ 8 s (s+1) \,
C_3\ {\rm Im} \, \bar{a}_3}
\label{eq:Gamma3_real}
\ee
with
\be
\boxed{\bar{a}_3 \ :=\   \frac{1}{6}\ \, \sum_{\md=-1}^1 \left[
a^{(\md)}_{2,1} + a^{(\md)}_{1,2}
\right].}
\nonumber
\ee
One could then determine ${\rm Im} \, \bar{a}_3$ by measuring $\Gamma_3$
in a weakly correlated regime, where the asymptotic behavior of the
three-body contact density $\Cr_3$ can be computed exactly.
A first option is the non-degenerate regime,
where we have computed $\Cr_3$ for negative or infinite scattering length.\footnote{For the homogeneous unpolarized unitary gas of density $n$,
we obtain $\Cr_3 \sim n^3 \, \left(\frac{\hbar^2}{m k_B T} \right)^{2-s}
\frac{\sqrt{243}\ \pi^3}{2^{2s+1}\,\Gamma(s+2)}$
in the non-degenerate limit [X.~Leyronas~and~F.~Werner, {\it ``Three-body contact for fermions. II. Non-degenerate limit''}, to be submitted].}
{\bl Other options are the weakly interacting regimes where $\ktyp\,|a_2|$ is small.\footnote{Although the three-body parameters depend on magnetic field, this dependence is smooth if no three-body resonance is crossed, and may be neglected in the vicinity of a given Feshbach resonance.}}


{\bl 
\subsection{Three-body contribution to the finite-range correction to the energy}
\label{sec:dE}

In this Section, we study the corrections to the zero-range model's energy coming from the finite range of the two-body interaction and/or an additional three-body interaction.
The stationary $N$-body Schr\"odinger equation is again given by (\ref{eq:schro_V3}).
We consider the case where there are no deeply bound dimers, so that the three-body parameters are real.
The zero-range model is approached in the zero-range regime where $b := {\rm Max}(b_2, b_3) \ll |a_2|, 1/\ktyp$.
In~particular, each eigenergy $E$ of (\ref{eq:schro_V3}) approaches a corresponding eigenenergy $E_{\rm ZRM}$ of the zero-range model.
\\We are interested in the energy difference between the finite-range and zero-range models,
\be
\delta\!E \ := \ E - E_{\rm ZRM}\,.
\nonumber
\ee
We find that  in the zero-range regime, 
\be
\delta\!E \ \simeq\ \delta\!E_2 \ +\ \delta\!E_3
\ee
plus higher-order corrections,
where $\delta\!E_2$ is given {\bl to leading order} by~\cite{WernerCastinRelationsFermions}
\begin{multline}
\delta\!E_2 \ =  \ 2\pi\  r_e \ N_\UP N_\down\ \int d^3\!c \,
\left( \prod_{i\neq1,3} d^3\!r_i \right)
A^*(\cc,\rr_2,\rr_4,\ldots,\rr_N)
\\ \left[
  E + \frac{\hbar^2}{4m}\,\Delta_\cc
  - 2 U(\cc)
  + \sum_{i\neq 1, 3} \left(\frac{\hbar^2}{2m}\,\Delta_{\rr_i}
  -U(\rr_i) \right) 
  \right] A(\cc,\rr_2,\rr_4,\ldots,\rr_N)
\label{eq:dE2}
\end{multline}
with $r_e$ the effective range of the two-body interaction $V_2$,
while $\delta\!E_3$ is given {\bl to leading order} by
\be
\boxed{\delta\!E_3 \ \simeq\ \  \frac{\hbar^2}{m}\ \ 4\,s\,(s+1)\ \ \sum_{\md=-1}^1 \left[
  C^{(\md)}_{2,1}\,  a^{(\md)}_{2,1}
\ +\   C^{(\md)}_{1,2}\,  a^{(\md)}_{1,2}
\right]}
\label{eq:dE_21_12}
\ee
in terms of the three-body contacts and three-body parameters.
If the two-body and three-body interaction potentials are rotationally invariant and $V_{2,1}=V_{1,2}$, then the six three-body parameters are all equal to a single $a_3$ and the expression simplifies to
\be
\boxed{\delta\!E_3 = \frac{\hbar^2}{m}\ 4\,s\,(s+1)\ \ 
C_3\,  a_3.}
\label{eq:dE3}
\ee

\vskip 0.12cm
\noindent \underline{\it Remarks:}
\bi

\item
$\delta\!E_2$ (resp. $\delta\!E_3$) comes from configurations where 2 (resp.~3) particles are close to each other.

  \item
  The relations (\ref{eq:dE_21_12},\ref{eq:dE3}) are reminiscent of the relation~\cite{TanLargeMomentum}
\be
\frac{d E}{d (-1/a_2)} \ = \ \frac{\hbar^2}{4\pi m}\ \,C_2
\label{eq:dE_da2}
\ee
which holds within the zero-range model.\footnote{\bl A relation which is more directly analogous to~(\ref{eq:dE_da2}) can be formulated in the mass-imbalanced case, see~(\ref{eq:dE_da3}).}

\item
 Typically, $a_3$ is of order $b^{2s}$, which gives
 $\delta\!E_3 \sim b^{2s}$.
 The latter scaling was already stated (for lattice models) in~\cite{KaplanTrap,KaplanXi} and was derived at the level of the third virial coefficient in~\cite{GaoEndoCastin}.

\item
  If $b_3 \lesssim b_2$,
  then $b_2\sim b$, and
since $\delta\!E_2$ is $\propto r_e$, which is typically of order $b_2$, we have
$|\delta\!E_3/ \delta\!E_2| \propto b^{2s-1}$.
Since $s>1/2$, we get $|\delta\!E_3/ \delta\!E_2| \ll 1$ in the zero-range regime, {\it i.e.} the~three-body correction to the energy is of higher order than the two-body correction.\footnote{\bl In the mass-imbalanced case discussed in App.~\ref{app:imbal}, the situation is reversed beyond a critical mass ratio, as was already noted in~\cite{GaoEndoCastin}.}$^,$\footnote{\bl Apart from the leading-order term (\ref{eq:dE2}) which is of order $b$, there are also higher-order contributions to $\delta\!E_2$, which we expect to contain no term of order $b^{2s}$ (since the three-body physics does not enter in $\delta\!E_2$) but only integer powers of~$b$. Accordingly, the $\propto b^{2s}$ contribution to $\delta\!E$ should be entirely given by (\ref{eq:dE_21_12}) or (\ref{eq:dE3}).}

\item Let us assume that (\ref{eq:dE3}) can be analytically continued to complex $a_3$, with $C_3$ still evaluated within the zero-range model. We then recover the expression (\ref{eq:Gamma3}) of the three-body loss rate, simply by substituting ${\rm Im}\, E = {\rm Im}\  \delta\!E_3$ into~(\ref{eq:ImE}), and using the fact that  {\bl $\Gamma=\Gamma_3$ and} $C_3\in\mathbb{R}$.
  Similarly, applying this procedure to (\ref{eq:dE_21_12}) yields $\Gamma_3$ in agreement with the sum of Eqs.~(\ref{eq:Gamma2,1_real},\ref{eq:Gamma1,2_real}).
  
\ei

To derive (\ref{eq:dE_21_12}), we consider a stationary state $\psi^{(0)}$
and the associated eigenenergy $E^{(0)}$
of the zero-range model, and the stationary state $\psi^{(1)}$ of the finite-range model whose energy $E^{(1)}$ is close to $E^{(0)}$ in the zero-range regime.
We are interested in $\delta\!E = E^{(1)} - E^{(0)}$.
\\Using the shorthand notations
\be
V(\rr_1,\ldots,\rr_N) \ := \  
  \sum_{\substack{i:\UP,\,j:\down}} V_2(\rr_{ij})
\ +\  \sum_{\substack{i<j:\UP,\,k:\down}} V_{2,1}(\RR_{ijk})
\ +\  \sum_{\substack{i<j:\down,\,k:\UP}} V_{1,2}(\RR_{ijk})
\nonumber
\ee
\be
H_0 \ := \ \sum_{i=1}^N \left[-\frac{\hbar^2}{2m}\ \Delta_{\rr_i} + U(\rr_i) \right]
\nonumber
\ee
we have
$(H_0 + V) \psi^{(1)} = E^{(1)} \psi^{(1)}$, while $\psi^{(0)}$ satisfies
$H_0 \psi^{(0)} = E^{(0)} \psi^{(0)}$ together with the two-body contact condition~(\ref{eq:BP}).
Let us consider
\be
\Delta \ := \ \la \psi^{(0)}, (H_0+V)\,\psi^{(1)} \ra
\ -\ 
\la H_0\,\psi^{(0)}, \psi^{(1)} \ra.
\nonumber \ee
We have $\Delta \ =\  \delta\!E\ \la \psi^{(0)} | \psi^{(1)} \ra$, and hence
$\delta\!E / \Delta\to 1$ in the zero-range limit.
\\To evaluate $\Delta$, we write it as
$\Delta = \Delta_0 + \ \la \psi^{(0)} | V | \psi^{(1)} \ra$
where
$\Delta_0 \ := \  \la \psi^{(0)}, H_0\,\psi^{(1)} \ra
-
\la H_0\,\psi^{(0)}, \psi^{(1)} \ra$.
We have $\Delta_0=0$,
assuming for simplicity that the interaction potentials ($V_2$, $V_{2,1}$ and $V_{1,2}$) are finite everywhere ({\it i.e.} excluding hard-wall potentials)
and do not diverge too quickly at short distances.\footnote{\bl Specifically, by using the divergence theorem as in Sec.~\ref{app:imbal,subsec:dE}, we find
the following sufficient conditions: $R^{2+s} \psi^{(1)}$ and $R^{3+s} \partial \psi^{(1)} / \partial R$ tend to zero when $R\to0$, and similarly,
$\tilde{R}^{2+s} \psi^{(1)}$ and $\tilde{R}^{3+s} \partial \psi^{(1)} / \partial \tilde{R}$ tend to zero when $\tilde{R}\to0$.}
Hence
\be
\Delta
\ \simeq \  \int \dr_1 \ldots \dr_N\ \left( \psi^{(0) *} \ V\  \psi^{(1)} \right)\!(\rr_1, \ldots , \rr_N).
\label{eq:DeltaV}
\ee
Since the potentials are short-ranged, the integral is dominated by the contributions from three regions, outside of which $V$ becomes negligible:
\bi
\item
   two particles of spins $\UP\down$ are nearby (at distance $\lesssim b_2$) 
   while all other interparticle distances are $\gg b$
 \item
   there is a triplet of particles of spins $\UP\UP\down$ which are nearby (their hyperradius is $\lesssim b$)
      while all interparticle distances other than the ones within that triplet are $\gg b$
 \item
   there is a triplet of particles of spins $\UP\down\down$ which are nearby (their hyperradius is $\lesssim b$)
      while all interparticle distances other than the ones within that triplet are $\gg b$.
\ei
Denoting the contributions of these regions by $\Delta_2\,$, $\Delta_{2,1}$ and $\Delta_{1,2}$ respectively, we thus have
$\Delta \simeq \Delta_2 + \Delta_{2,1} + \Delta_{1,2}$.
The two-nearby-particle contribution $\Delta_2$ is given by the r.h.s. of~(\ref{eq:dE2}), as shown in Appendix~\ref{app:dE2}, in agreement with~\cite{WernerCastinRelationsFermions}.
It remains to evaluate the contribution $\Delta_{2,1}$ coming from three nearby particles of spins $\UP\UP\down$. We introduce a length
$d_3$
{\bl satisfying (\ref{eq:<<d},\ref{eq:d<<},\ref{eq:l_2}).}
$\Delta_{2,1}$ is then given by the contribution to the integral (\ref{eq:DeltaV}) coming from the region of $(\rr_1,\ldots,\rr_N)$ such that there is a triplet of particles $(i,j,k)$ of spins $(\UP,\UP,\down)$ of hyperradius  $R_{ijk} <
d_3$
while all interparticle distances other than the ones within the triplet $(i,j,k)$ are
$\gg d_3$
(the result does not depend on the value of
$d_3$
within the range (\ref{eq:<<d},\ref{eq:d<<}), as we will see).
In the region where these conditions are met for the triplet $(1,2,3)$,
{\bl {\it i.e.} when (\ref{eq:R<d},\ref{eq:d<<rij}) hold,}
we expect a factorization
\be
\psi^{(1)}(\rr_1,\ldots,\rr_N)
\simeq
\sum_{\md=-1}^{+1} \Psi_\md(\RR)\ 
 B_\md^{(1)}(\CC;\rr_4,\ldots,\rr_N),
\label{eq:psi1_R0}
 \ee
 as well as [cf.~(\ref{eq:R0})]
 \be
\psi^{(0)}(\rr_1,\ldots,\rr_N)
\simeq
\sum_{\md=-1}^{+1} \Psi^{(0)}_\md(\RR)\ 
B_\md(\CC;\rr_4,\ldots,\rr_N)
\nonumber
 \ee
 with $\Psi^{(0)}_\md(\RR) \ := \ R^{s-2}\ \phi_\md(\Oo)$.
 Furthermore, we can approximate $B^{(1)}_\md$ by $B_\md$ in (\ref{eq:psi1_R0}), given that we are in the zero-range regime and all distances between the points $\CC,\rr_4,\ldots,\rr_N$ are $\gg b$.
Also using the fact that each triplet gives the same contribution by fermionic antisymmetry,
we get
\be
\Delta_{2,1} \simeq
\frac{N_\UP (N_\UP-1) N_\down}{2} \frac{3\sqrt{3}}{8}\sum_{\md,\md'=-1}^{1}
W_{\md,\md'} \int d^3\!C\,d^3\!r_4 \ldots d^3\!r_N\ \left( B_\md^{\,*}\,B_{\md'}^{(1)} \right)\!\!(\CC\,;\,\rr_4,\ldots,\rr_N)
\label{eq:Delta21}
\ee
where
\be
W_{\md,\md'} \ := \
\int_{R<d_3}
d^6R\ \left(
  \Psi_\md^{(0)\,*}\,\Vr\,\Psi_{\md'}
  \right)\!\!(\RR)
\nonumber
\ee
and $\Vr(\RR)$ was defined in (\ref{eq:def_Vr}).
To evaluate $W_{\md,\md'}$, we first use the Schr\"odinger equation~(\ref{eq:schro3_V3}) to replace $\Vr$ by $\frac{\hbar^2}{m} \Delta_\RR$.
Since we also have $\Delta_\RR\,\Psi^{(0)}=0$, we can write
$W_{\md,\md'} \ = \ \frac{\hbar^2}{m}\
\int_{R<d_3}
\ d^6\!R\
\left[
  \Psi_\md^{(0)\,*}\,\Delta_\RR\,\Psi_{\md'}
  -
\Psi_{\md'}  \,\Delta_\RR\,  \Psi_\md^{(0)\,*}
  \right]\!\!(\RR)$.
Rewriting the integrand as $\gr_\RR \cdot \left[
  \Psi_\md^{(0)\,*}\,\gr_\RR\,\Psi_{\md'}
  -
\Psi_{\md'}  \,\gr_\RR\,  \Psi_\md^{(0)\,*}
\right]$ and
applying the divergence theorem yields
\be W_{\md,\md'} \ = \ \frac{\hbar^2}{m}\ d^5\ \int d^5\Omega \ \left[ \Psi_\md^{(0)\,*}\,\partial_R \Psi_{\md'} \, - \, \Psi_{\md'}\, \partial_R \Psi_\md^{(0)\,*}
  \right]_{R=d_3}\,.
\nonumber \ee
We can then use the asymptotic behavior (\ref{eq:a21}) of $\Psi_\md$
(indeed, for $R=
d_3
\gg b$, {\bl all three interparticle distance are $\gg b$} except in a small region of hyperangles). 
This gives
$W_{\md,\md'} \ = \ \frac{\hbar^2}{m}\ 2\,s\ \,\delta_{\md,\md'}\  a_{2,1}^{(\md)}$.
Substituting this into~(\ref{eq:Delta21}) and using~(\ref{eq:C3_BB}) yields
$\Delta_{2,1} \ \simeq\ \  \frac{\hbar^2}{m}\ \ 2\,s\,(s+1)\ \ \sum_{\md=-1}^1  C^{(\md)}_{2,1}\,  a^{(\md)}_{2,1}$.
\\The same reasoning gives an analogous expression for $\Delta_{1,2}$,
which yields the final expression~(\ref{eq:dE_21_12}) for $\delta\!E_3 \simeq \Delta_{2,1} + \Delta_{1,2}$.
}

{\bl
\section{Summary and outlook}

We have shown that the three-body contact $C_3$ is a useful concept for the fermionic $N$-body problem with resonant interactions, in the standard regime of mass ratio where there is no Efimov effect.
Within the zero-range model, the three-body contact controls the number of nearby triplets, the third order density correlation function at short distances, the tail of the center-of-mass momentum distribution of nearby pairs, {\gre and the tail of the two-particle momentum distribution}; $C_3$ also has a simple expression in terms of the $N$-body wavefunction in the limit where three particles are nearby.
Beyond the zero-range model, for a small finite interaction range, we introduced a small three-body parameter $a_3$, and we showed that the formation rate $\Gamma_3$ of deeply bound dimers by three-body recombination equals $C_3\ {\rm Im}\,a_3$ times an explicit prefactor;
we also showed that the finite-range correction to the energy has a contribution equal to $-\frac{\hbar}{2}\,C_3\,a_3$ times the same prefactor.
{\brown With respect to the relation between $\Gamma_3$ and the number of nearby triplets stated in~\cite{Petrov4body2004},} {\brown the present work adds a derivation, and an expression of the prefactor in terms of ${\rm Im}\,a_3$.}


{\brown Furthermore, we considered the general case where} there are two different contributions $C_{2,1}$ and $C_{1,2}$ to the three-body contact,
corresponding to the spin configurations $\UP\UP\down$ and $\UP\down\down$ for the associated three-body problem, {\brown which} can be further broken up into the contributions
$C_{2,1}^{(\md)}$ and $C_{1,2}^{(\md)}$
for each value $\md \in \{ -1, 0, 1 \}$ of the angular-momentum quantum number of the three-body problem.
Most relations only involve $C_3$, $C_{2,1}$ or $C_{1,2}$.
The only relations involving $C_{2,1}^{(\md)}$ and $C_{1,2}^{(\md)}$
are the ones for finite-range interactions,
in the general case where interactions are not invariant under rotation and under exchange between $\UP$ and $\down$.
In this case, the three-body parameter also depends on the spin and angular momentum indices.

Nevertheless, for a homogeneous unpolarized gas, 
$\Gamma_3$ simply equals $C_3\,{\rm Im}\,\bar{a}_3$ times an explicit prefactor, with $\bar{a}_3$ the mean three-body parameter.
For the unitary gas in the non-degenerate regime, we announced the result of our computation of $C_3$ (see footnote 21), which would allow one to determine ${\rm Im}\,\bar{a}_3$ by measuring $\Gamma_3$.
Measuring $\Gamma_3$ in the low-temperature regime would then
allow one to experimentally test the power-law $\Cr_3 = \zeta_3\,n^{(2s+5)/3}$ and determine the associated many-body parameter $\zeta_3$ 
whose computation is an open theoretical challenge.
}

\longthanks
We are grateful  
to S.~Tan,
C.~De~Daniloff and F.~Chevy for particularly
important input.
We~also thank Y.~Castin,  N.~Navon, D.~Petrov, T.~Yefsah and
the  {\it ultracold Fermi gases} team at LKB for stimulating discussions.
F.W.
acknowledges the hospitality of the Aspen Center for Physics, and
 support from
ERC (project {\it Critisup2}, H2020 Adv-743159)
and ANR (project {\it LODIS}, ANR-21-CE30-0033).
\appendix

\section{The unitary three-body problem}
 \label{app:unit_hyp}

In this Appendix we review, for self-containedness of the article, the known solution of the three-body problem at the unitary limit~\cite{Efimov,Efimov73}
(see also Refs.~\cite{Werner3corpsPRL,WernerThese}).
Consider an eigenstate $\psi(\rr_1,\rr_2,\rr_3)$ of the three-body problem in free space.
With the convention that particles $(1,2,3)$ have spins $(\UP,\UP,\down)$,
the fermionic antisymmetry reads
$\psi(\rr_1,\rr_2,\rr_3) = - \psi(\rr_2,\rr_1,\rr_3)$.
Restricting to zero center-of-mass momentum, $\psi$ only depends on the relative 
positions between the three particles, 
which can be parameterized by 
the Jacobi coordinates
\bea
\rr &=& \rr_3-\rr_1
\nonumber 
\\ \frac{\sqrt{3}}{2} \rrho &=& \rr_2 - \frac{\rr_1+\rr_3}{2}.
\label{eq:def_jaco}
\eea
It is convenient to introduce
the six-dimensional vector 
\be
\RR=(\rr,\rrho)
\label{eq:def_RR}
\ee
whose norm
$R=\| \RR \|=\sqrt{r^2+\rho^2}$
is the hyperradius,
while its direction $\RR/R$ can be 
parameterized by five hyperangles denoted collectively by $\Oo$,
\be
\Oo \ \Longleftrightarrow\  \RR/R.
\label{eq:def_Oo}
\ee 
The three-body Schr\"odinger equation then writes
\be
-\frac{\hbar^2}{m} \Delta_{\RR} \psi(\RR) = E \, \psi(\RR)
\label{eq:schro_3b}
\ee
with the contact condition:
there exists $\Ar$ such that
\be
\psi(\RR) \underset{r\to0}{=}\  \left( \frac{1}{r} - \frac{1}{a_2} \right)\ \Ar(\rrho) + O(r)
\label{eq:BP_3corps}
\ee
(the contact condition between particles 2 and 3 then automatically holds by antisymmetry).

At the unitary limit $a_2=\infty$, this contact condition becomes scale invariant.
Accordingly, it only acts on the hyperangles;
explicitly, it can be expressed as
$\frac{\partial}{\partial \alpha} \left( \sin \alpha\ \, \psi \right)_{\alpha=0} = 0$
where $\alpha={\rm arctan}(r/\rho)$.
As a result, the unitary three-body problem is separable in hyperspherical coordinates:
One can look for eigenstates of the factorized form
\be
\psi(\RR) = \frac{F(R)}{R^2} \, \phi(\Oo)
\nonumber
\ee
(where the factor $1/R^2$ is introduced for later convenience),
and the three-body problem~(\ref{eq:schro_3b},\ref{eq:BP_3corps}) 
separates into
\bi

\item
a hyperradial problem
\be
\left( -\frac{d^2}{dR^2} - \frac{1}{R} \frac{d}{dR} + \frac{s^2}{R^2} \right) F(R) = 
\frac{m}{\hbar^2}\, E\,
F(R)
\label{eq:schro_R}
\ee

\item
a hyperangular problem, defined by
\be
T_\Oo \, \phi(\Oo) \ =\ -s^2 \phi(\Oo)
\label{eq:eigen_hyperang}
\ee
together with the contact condition
\be
\frac{\partial}{\partial \alpha} \Big[ \sin \alpha\ \phi(\Oo) \Big]_{\alpha=0} = 0
\label{eq:cc_ang}
\ee
and the antisymmetry constraint.
\\Here $T_\Oo$ is a differential operator acting on the hyperangles,
defined by
\be
\Delta_{\RR} = \frac{1}{R^2} \left(\frac{\partial^2}{\partial R^2} + \frac{1}{R} \frac{\partial}{\partial R} + \frac{1}{R^2}\, T_{\Oo} 
\right) R^2\,.
\label{eq:def_TOm}
\ee

\ei

There is a discrete spectrum of values for $s^2$, which are all real and positive in the present case of equal-mass fermions, so that we can take $s$ real and positive;
we denote the set of allowed $s$ by
$\{ s_\nu \}$ where $\nu$ is a discrete index.
The corresponding hyperangular eigenfunctions $\{ \phi_\nu(\Oo) \}$ form an orthonormal basis for the hyperangular scalar product
\be
(f|g) \equiv \int f(\Oo)^* \, g(\Oo) \, d^5\Omega
\ee
where $ d^5\Omega$ denotes the differential solid angle in six-dimensional space,
$d^6\!R = d^5\Omega \ R^5 \,dR\,$;
this can be deduced from 
the self-adjointness of the unitary three-body problem in an isotropic harmonic trap 
and can also be checked by explicit analytical calculations in the $l=0$ subspace~\cite{WernerTheseCompleteness}
(for a mathematical proof of self-adjointness in free space, see Refs.~\cite{MinlosLNP,MinlosShermatov,TetaEPL}).


The hyperangular problem (\ref{eq:eigen_hyperang},\ref{eq:cc_ang}) is analytically solvable, {\bl with two types of solutions.
The first type are common solutions of the unitary and non-interacting hyperangular problems, whose wavefunction $\phi(\Oo)$ vanishes when two particles approach each other ({\it i.e.} for $\alpha\to0$);
the corresponding  $s$ take  integer values.
The second type are the following truly interacting solutions:}
\bi
\item
For each $l$, let us denote by $\{s_{l,n}; n\in\mathbb{N} \}$ the allowed values of $s$, in increasing order ($s_{l,0} < s_{l,1} < \ldots$). 
The index $\nu$ can be identified with the set of quantum numbers $(l,\md,n)$.
All the eigenfunctions $\phi_{(l,\md,n)}$ with $-l\leq \md \leq l$ correspond to the same $s_{l,n}$,
so that each $s_{l,n}$ is $2l+1$ times degenerate.
For $l=1$, the $s_{l,n}$ are the real positive solutions different from 1 of the transcendental equation~(\ref{eq:transc_s})
[the solution $s=1$ should be discarded since it leads to an identically vanishing $\varphi(\alpha)$].
For arbitrary $l$, the $s_{l,n}$ also solve transcendental equations (see, {\it e.g.}, \cite{LeChapitreIn2} and refs. therein);
the smallest of all $s_{l,n}$ is $s_{l=1, n=0}$, which is denoted by $s$ for short throughout the article.

\item
The eigenfunctions are
\be
\phi(\Oo) = \Nr (1-\hat{P}) \ \frac{\varphi(\alpha)}{\sin(2\alpha)}\ Y_l^\md({\hat{\bm \rho}})
\label{eq:phi}
\ee
where $l\in\mathbb{N}$ and $\md\in\{-l;\ldots;l\}$ are the total-angular-momentum quantum numbers,\footnote{More explicitly, the considered wavefunction is an eigenstate of $\LL^2$ with eigenvalue $l(l+1)\hbar^2$, and of $L_z$ with eigenvalue $\md\hbar$, where $\LL$ is the relative angular momentum of the 3 particles, $\LL=-i\hbar(\rr \times \gr_\rr + \rrho \times \gr_{\rrho} )$. Note that the total angular momentum of the three particles
is the sum of their relative and center-of-mass angular momenta:
$-i \hbar\sum_{j=1}^{3} \rr_j \times \gr_{\rr_j} = \LL -i\hbar\,\CC \times \gr_\CC$.
}
whereas
$\Nr>0$ is a normalization constant such that $(\phi|\phi)=1$.
Note that $\Nr$ does not depend on $\md$, as follows from the relation $L_{\pm}\ \tilde{\phi}_\md = \hbar \sqrt{ l (l+1) - \md (\md+1) }\ \tilde{\phi}_{\md\pm1}$ where $L_{\pm} := L_x \pm i L_y$ and $\tilde{\phi}_\md := (1-\hat{P}) \ \frac{\varphi(\alpha)}{\sin(2\alpha)}\ Y_l^\md(\hat{\bm \rho})$.
\\For $l=1$, 
  \be
\varphi(\alpha) = -s\,\cos\!\left[s\left(\frac{\pi}{2}-\alpha\right)\right] 
 + \sin\!\left[s\left(\frac{\pi}{2}-\alpha\right)\right]\, \tan \alpha
\label{eq:varphi_l=1}
\ee
(for arbitrary $l$, see, {\it e.g.},~\cite{LeChapitreIn2} and refs. therein).
We refer to the $\{ \phi_\nu(\Oo) \}$ as 
unitary hyperangular wavefunctions.\footnote{Like the standard hyperspherical harmonics,
the $\phi_\nu$ are eigenstates of the Laplacian on the hypersphere, Eq.~(\ref{eq:eigen_hyperang}); but they also satisfy
the unitary-limit contact condition Eq.~(\ref{eq:cc_ang})
(together with fermionic antisymmetry)
which leads to non-integer
eigenvalues $s^2$.}
We use the shorthand notation $\phi_\md := \phi_{(l=1,\md,n=0)}$.

\ei

{\bl The value of the normalization constant, for $l=1$, is
  \be
  \Nr = 0.31149\ldots
  \label{eq:Nr}
  \ee
We computed this value as follows.
  We have $\mathcal{N}=\sqrt{3/J(s)}$ with
  $J(s)=\sum_{\md=-1}^1(\tilde{\phi}_{\md}|\tilde{\phi}_{\md})$.
\\We evaluate the sum over $\md$
thanks to the identity:
$\sum_{\md=-1}^{1}Y_1^{\md}({\bf \hat{u}})^*\,Y_1^{\md}({\bf \hat{u}'})=\frac{3}{4\,\pi}P_1({\bf \hat{u}}\cdot{\bf \hat{u}'})$ for any unit vectors ${\bf \hat{u}}$ and ${\bf \hat{u}'}$. This yields
\be
J(s)=\frac{3}{4\pi}\,
\int d^5\Omega\,\left[
\left(\frac{\varphi(\alpha)}{\sin(2\alpha)}\right)^2
+
\left(\frac{\varphi(\alpha')}{\sin(2\alpha')}\right)^2
-2\,P_1\!\left(\hat{\bm \rho}\cdot{\bf \hat{\bm \rho}'}
\right)\frac{\varphi(\alpha)}{\sin(2\alpha)}\,\frac{\varphi(\alpha')}{\sin(2\alpha')}
\right]\label{eq:Js}
\ee
where $\alpha'$ and ${\bf \hat{\bm \rho}'}$ are 
obtained from $\alpha$ and $\hat{\bm \rho}$ by permutation of particles 1 and~2.
The integrand 
in (\ref{eq:Js})
only depends on $\alpha$ and $\alpha'$, since 
$\hat{\bm \rho}\cdot\hat{\bm \rho}'=\frac{\sin^2(\alpha)+\sin^2(\alpha')-5/4}{\cos(\alpha)\,\cos(\alpha')}$.
Moreover $\alpha'$ is a function of $\alpha$ and of $u=\hat{\bf r}\cdot\hat{\bm \rho}$, since
$u=\frac{2}{\sqrt{3}}\times\frac{3/4-\sin^2(\alpha)/2-\sin^2(\alpha')}{\cos(\alpha)\sin(\alpha)}$.
Hence the integrand in (\ref{eq:Js})
 is a function of $\alpha$ and $u$. 
To evaluate the integral we use the formula 
$\int d^5\Omega\
f(\Oo)
=
\int_0^{\pi/2}d\!\alpha\ \sin^2(\alpha) \, \cos^2(\alpha)
\int d\!\hat{\bf r} \, d\!\hat{\bm \rho}\, f(\alpha,\hat{\bf r},\hat{\bm \rho})$.
Since the integrand is independent of $\hat{\bf r}$ and of the azimuthal angle of $\hat{\bm \rho}$ w.r.t. $\hat{\bf r}$,
we can integrate over them,
 which just gives a factor $4\pi\times2\pi$. 
 We are left with the integral over $\alpha$ and $u$. 
 The change of variable $u \longrightarrow \alpha'$ then yields
\begin{multline}
J(s)=8\sqrt{3}\pi\int_{\mathcal{D}}d\!\alpha\,d\!\alpha'\,
\cos(\alpha)\sin(\alpha)\cos(\alpha')\sin(\alpha')\\
\times\left[
\left(\frac{\varphi(\alpha)}{\sin(2\alpha)}\right)^2
+
\left(\frac{\varphi(\alpha')}{\sin(2\alpha')}\right)^2
-2\,P_1\left(
\frac{\sin(\alpha)^2+\sin(\alpha')^2-\frac{5}{4}}{\cos(\alpha)\cos(\alpha')}
\right)\frac{\varphi(\alpha)}{\sin(2\alpha)}\frac{\varphi(\alpha')}{\sin(2\alpha')}
\right]\label{eq:Js2}
\end{multline}
where $\mathcal{D}$ is the domain
$\left\{
\left(\alpha,\alpha'\right) \, \Big|\,\frac{\pi}{3}\,\leq\alpha+\alpha'\leq\,\frac{2\pi}{3},\,
|\alpha-\alpha'|\leq\,\frac{\pi}{3}
\right\}$.
 We evaluate analytically the integrals over $\alpha$ and $\alpha'$  for the first and second term of (\ref{eq:Js2});
  for the third term, we  evaluate analytically the integral over $\alpha'$, and perform numerically the remaining integration over $\alpha$.


  }

\section{Homogeneous gas} \label{app:C3m}


Let us show that for the homogeneous gas at equilibrium,
$C^{(\md)}_{2,1}$
is independent of $\md$.
We will use the following lemma:
Let $\hat{\Or}$ be an operator acting on  functions of $\Oo$,
such that
\be
(\hat{\Or}\,\phi_\md)(\Oo) = \lambda_\md\,\phi_\md(\Oo).
\nonumber
\ee
Then,
\be
  \frac{N_\UP\,(N_\UP-1)\,N_\down}{2}\ \left<
\hat{\Or}
\ \theta(\epsilon-R) \right> \underset{\epsilon\to0}{\sim}\ \epsilon^{2s+2} \sum_{\md=-1}^1 \lambda_\md\ C^{(\md)}_{2,1}
\ee
with $\theta$ the Heaviside function.
This follows from Eq.~(\ref{eq:R0}) and the fact that the $\phi_\md(\Oo)$ are orthonormal.

We then consider $Q := \langle \hat{L}_z\ \,\theta(\epsilon-R) \rangle$.
In the absence of time-reversal symmetry breaking, we have $Q=0$.
On the other hand, applying the above lemma to $\hat{\Or}=\hat{L}_z$ gives $Q \propto C^{(1)}_{2,1} - C^{(-1)}_{2,1}$. Hence $C^{(1)}_{2,1} = C^{(-1)}_{2,1}$.

Since the system is isotropic, the quantity $\langle \hat{L}_i^{\phantom{i}2}\,\theta(R-\epsilon) \rangle$ is the same for $i=x, y$ and $z$;
hence
$\langle \hat{L}_z^{\phantom{i}2}\,\theta(R-\epsilon) \rangle
= \langle \hat{\bf L}^{2}\,\theta(R-\epsilon) \rangle / 3$.
Applying the lemma to both sides then yields
$C^{(1)}_{2,1} + C^{(-1)}_{2,1} = 2 \sum_{\md=-1}^1 C^{(\md)}_{2,1} / 3$,
which gives $C^{(1)}_{2,1} = C^{(0)}_{2,1}$.

Finally we note that for the unpolarized gas ($N_\UP = N_\down$) at equilibrium, the two states $\UP$ and $\down$ play a symmetric role, so that $C^{(\md)}_{2,1} = C^{(\md)}_{1,2}$.

{\bl 
\section{Two-body contribution to the energy correction}  \label{app:dE2}
In this Appendix we show that $\Delta_2$ is given by the r.h.s. of~(\ref{eq:dE2}), to leading order in the zero-range regime. In the mass-balanced case, to leading order, we have $\delta\!E \simeq \Delta_2$, so that we recover the result of~\cite{WernerCastinRelationsFermions}.
We provide the present derivation because it is not identical (although similar) to the one in Ref.~\cite{WernerCastinRelationsFermions} (the quantity $\Delta_2$, as defined here, does not explicitly appear in~\cite{WernerCastinRelationsFermions}).

Let us introduce a length $d$ such that
\be
b \ll d \ll 1/\ktyp\,.
\label{eq:range_d2}
\ee
$\Delta_2$ is given by the contribution to the integral~(\ref{eq:DeltaV}) coming from the region of $(\rr_1,\ldots,\rr_N)$ such that there is a pair of particles $(i,j)$ of spins $(\UP,\down)$ separated by a distance  $r_{ij} < d$ while all interparticle distances are $\gg d$
[the result will not depend on the value of $d$ within the range (\ref{eq:range_d2})].
\\Let us denote by $\Rr_{13}$ the region where these conditions are met for the pair $(1,3)$,
\be \Rr_{13} := \left\{ (\rr_1,\ldots,\rr_N) \ {\rm such\ that}\ \  r\equiv r_{13}<d \ {\rm and\ all\ other\ } r_{ij} \gg d \right\}.\nonumber \ee
Since all $\UP\down$ pairs of particles contribute equally,
\be
\Delta_2
\ = \  N_\UP\,N_\down\ \int_{\Rr_{13}} \dr_1 \ldots \dr_N\ \left( \psi^{(0) *} \ V\  \psi^{(1)} \right)\!(\rr_1, \ldots , \rr_N).
\label{eq:DeltaV13}
\ee

In $\Rr_{13}$, we expect a factorization of the many-body wavefunction of the finite-range model
\be
\psi^{(1)}(\rr_1, \ldots , \rr_N) \simeq \chi_\Er(r) \ A(\cc; \rr_2, \rr_4, \ldots , \rr_N)
\label{eq:psi1_r13}
\ee
where we can assume $\chi_\Er$ to be rotationally invariant, as was checked in~\cite{WernerCastinRelationsFermions}.
\\Injecting the ansatz (\ref{eq:psi1_r13}) into the $N$-body Schr\"odinger equation (\ref{eq:schro_V3}) yields, in the region $\Rr_{13}$,
\be
-\frac{\hbar^2}{m} \Delta_\rr\chi_\Er(r) + V_2(r)\,\chi_\Er(r) \ = \ \Er \, \chi_\Er(r)
\label{eq:chi}
\ee
with
\be
\Er :=  E - 2\,U(\cc) - \sum_{k\in\{2; 4; \ldots ; N\}} U(\rr_k)
+ \frac{1}{A(\cc; \rr_2, \rr_4, \ldots , \rr_N)}\,\left(
\frac{\hbar^2}{4m}\Delta_\cc
+ \frac{\hbar^2}{2m}\sum_{k\in\{2; 4; \ldots ; N\}}\Delta_{\rr_k}
  \right)\,A(\cc; \rr_2, \rr_4, \ldots , \rr_N)\,, \nonumber
  \ee
  up to corrections that are negligible in the zero-range regime, as was checked in~\cite{WernerCastinRelationsFermions}.
  We omitted the dependence of $\Er$ on $(\cc; \rr_2, \rr_4, \ldots , \rr_N)$ to alleviate notations.
  We neglected all interaction potentials other than $V_2(r)$, because we are far outside their ranges: In the considered region $\Rr_{13}$, all interparticle distances other than $r \equiv r_{13}$, and hence also all triplet hyperradii, are $\gg b$. For the same reason, we will replace $V$ by $V_2$ in (\ref{eq:DeltaV13}).

  For the zero-range model, we also expect a factorization of the many-body wavefunction in~$\Rr_{13}$,
  \be
\psi^{(0)}(\rr_1, \ldots , \rr_N) \simeq \chi^{(0)}_\Er(r)  \ A(\cc; \rr_2, \rr_4, \ldots , \rr_N)
\label{eq:psi0_r13}
\ee
where $\chi^{(0)}_\Er(r)$ is the $s$-wave two-body scattering state at energy $\Er$ for the zero-range model, \\{\it i.e.}~the solution of
\be
-\frac{\hbar^2}{m} \Delta_\rr\chi^{(0)}_\Er(r) = \Er \, \chi^{(0)}_\Er(r)
\label{eq:chi0}
\ee
with the contact condition: $\chi^{(0)}_\Er(r) = 1/r - 1/a_2 + O(r)$ for $r\to0$.
Here we normalized $\chi^{(0)}_\Er$ in such a way that the function $A$ in (\ref{eq:psi0_r13}) is the same one than in (\ref{eq:BP}).
The solution is
\be
r\,\chi^{(0)}_\Er(r) = \frac{1}{f_k^{(0)}}\,\frac{\sin(kr)}{k} + e^{ikr}
\label{eq:chi0_sol}
\ee
where $f_k^{(0)} = -(1/a_2 + i k)^{-1}$ is the scattering amplitude of the zero-range model, and $k := \sqrt{m \Er} / \hbar$, with the determination
$k = i \sqrt{- m \Er} / \hbar$ if $\Er<0$.
\\We note that for $r\to0$, from the Taylor expansion of (\ref{eq:chi0_sol}), we get a subleading singular contribution to the asymptotic expansion of $\psi^{(0)}$ given by $-k^2 r A / 2$, in agreement with Eqs.~(135,136) of~\cite{WernerCastinRelationsFermions}.
From this we can infer that $k$ is typically $\lesssim \ktyp$.

Scattering theory gives the large-distance behavior of the finite-range scattering state $\chi_\Er$:
\be
r\,\chi_\Er(r)
\underset{r\gg b_2}{\simeq}\ \frac{1}{f_k}\,\frac{\sin(kr)}{r} + e^{ikr}
\label{eq:chi_inf}
\ee
where $f_k$ is the $s$-wave scattering amplitude associated to the interaction potential $V_2(r)$.
\\Here we normalized $\chi_\Er$ in such a way that $\chi_\Er(r) \sim 1/r$ for $\Er\to0$ and $r\to\infty$, in agreement with the assumption that the same function $A$ appears in (\ref{eq:psi0_r13}) and (\ref{eq:psi1_r13}).

Next, in (\ref{eq:DeltaV13}), we can thus replace $V$ by $V_2$
and substitute (\ref{eq:psi0_r13},\ref{eq:psi1_r13}), which yields
\be
\Delta_2
\ \simeq \  N_\UP\,N_\down\ \int \dc \, \dr_2 \, \dr_4 \, \ldots \, \dr_N\ \ \left| A(\cc; \rr_2, \rr_4, \ldots , \rr_N) \right|^2 \ W(\cc; \rr_2, \rr_4, \ldots , \rr_N)
\label{eq:dE2_W}
\ee
where
\be
W \ := \ \int_{r<d} \, \dr \ \left(\chi^{(0)}_\Er V_2 \chi_\Er \right)\!(r)\,. \nonumber
\ee
It remains to evaluate $W$. Using (\ref{eq:chi},\ref{eq:chi0}), we rewrite it as
$W = \frac{\hbar^2}{m} \int_{r<d} \ \dr \ \left(\chi^{(0)}_\Er \Delta_\rr \chi_\Er
- \chi_\Er \Delta_\rr \chi^{(0)}_\Er \right)\!(r)$.
This gives, by the divergence theorem,
$W = \frac{4\pi\hbar^2}{m}\, \Wr$ with
$\Wr := \left(\chi^{(0)}_\Er \frac{\partial \chi_\Er}{\partial r}
- \chi_\Er \frac{\partial \chi^{(0)}_\Er}{\partial r} \right)_{r=d}$.
\\Equivalently,
$\Wr = \left(r\chi^{(0)}_\Er \frac{\partial (r\chi_\Er)}{\partial r}
- r\chi_\Er \frac{\partial (r\chi^{(0)}_\Er)}{\partial r} \right)_{r=d}$;
since $d\gg b_2$, we can directly substitute (\ref{eq:chi0_sol},\ref{eq:chi_inf}) and their derivatives, which yields
$\Wr \simeq 1/f_k - 1/f^{(0)}_k$.
\\Since $k b_2 \lesssim \ktyp b_2 \ll 1$, we can
use the low-energy expansion of the scattering amplitude:
$-1/f_k = 1/a_2 + i k - k^2 r_e / 2 + \ldots$ in the limit $k b_2\to 0$,
where $r_e$ is by definition the effective range. This gives
$W \simeq 2\pi\Er\,r_e$.
Substituting this into (\ref{eq:dE2_W}), we conclude that $\Delta_2$ is indeed given by the r.h.s. of~(\ref{eq:dE2}).
}

{\bl
  
\section{Mass-imbalanced case} \label{app:imbal}

In this Appendix we consider
the case where
the $\UP$ and $\down$ fermions have different masses, $m_\UP > m_\down$ for definiteness.
Experimentally, this is realized in a mixture of two fermionic species,
such as $^{40}$K-$^6$Li~\cite{GrimmUltrafast}, $^{161}$Dy-$^{40}$K~\cite{GrimmDyK}, or  $^{53}$Cr-$^6$Li~\cite{ZaccantiLiCrFeshbach}.
In Section~\ref{app:sec:gen}, we extend the relations obtained in the main text to the mass-imbalanced case.
In Section~\ref{app:sec:simpl} we show that,
when the mass ratio exceeds a critical value,
a~conceptual simplification takes place:
A generalized zero-range model can be introduced,
within which
relations involving the three-body parameters
can be formulated and derived more directly.
{\gre The results of this Appendix are valid if $m_\UP/m_\down$ is smaller than the threshold where the five-body Efimov effect appears~\cite{PetrovBazak5body}, which implies that the four-body Efimov effect~\cite{CMP} and three-body Efimov effect~\cite{Efimov73} do not take place either;
  moreover $m_\UP/m_\down$ should not be too close to the four-body and five-body Efimov thresholds, as discussed in Section~\ref{sec:max_ratio}.}

Obviously, in the $N$-body Schr\"odinger equation, the mass $m$ is replaced by 
 the mass $m_i$ of particle $i$ ($m_i = m_\UP$ or $m_\down$ depending on the spin of particle $i$):
\be
\sum_{i=1}^N \left[-\frac{\hbar^2}{2m_i}\Delta_{\rr_i} + U(\rr_i) \right] \psi = E\,\psi
\label{eq:schro_imbal}
\ee
for the zero-range model, and
\be
\sum_{i=1}^N \left[-\frac{\hbar^2}{2m_i}\Delta_{\rr_i} + U(\rr_i) \right] \psi
\ +\  \sum_{\substack{i:\UP,\,j:\down}} V_2(r_{ij}) \ \psi
 = E\,\psi,
\label{eq:schro_V2_imbal}
\ee
for the finite-range model.
\\It will prove convenient to
introduce the
angles $\varphi$ and $\tilde{\varphi}$
related to the mass ratio by
\be
   {\rm sin}\ \varphi \ = \ \frac{m_\UP}{m_\UP + m_\down}\ ,  \nonumber
   \ \ \ \ \ \ \ \ \  \ \ \ \ \ \ \ \ \   {\rm sin}\ \tilde{\varphi} \ = \ \frac{m_\down}{m_\UP + m_\down}. \nonumber
   \ee
The definitions of the center-of-mass and Jacobi coordinates should be generalized to the unequal-mass case.
The center-of-mass of particles 1 and 3,
which appears in the two-body contact condition~(\ref{eq:BP}),
is now $\cc = (m_\UP \rr_1 + m_\down \rr_3)/(m_\UP+m_\down)$.
The center-of-mass of particles 1,2,3 (assumed to have spins $\UP,\UP,\down$) is now 
$\CC = [m_\UP (\rr_1 + \rr_2) + m_\down \rr_3]/(2 m_\UP + m_\down)$,
and the Jacobi coordinate $\rrho$ is now
$(\rr_2 - \cc) / {\rm cos}\, \varphi$,
while we still have $\rr := \rr_3 - \rr_1$ and $\RR := (\rr,\rrho)$.
Similarly, the center-of-mass of particles 1,3,4 (of spins $\UP,\down,\down$) is now 
$\tilde{\CC} = [m_\UP \rr_1  + m_\down (\rr_3+\rr_4)]/(m_\UP + 2 m_\down)$,
and the Jacobi coordinate $\tilde{\rrho}$ is now
$(\rr_4 - \cc) / {\rm cos}\, \tilde{\varphi}\ $,
while we still have $\tilde{\rr} := \rr_1 - \rr_2$ and $\tilde{\RR} := (\tilde{\rr},\tilde{\rrho})$.
The three-body Schr\"odinger equation is then still given by (\ref{eq:schro_3b}) provided we define $m$ as twice the reduced mass,
\be
m \ :=\  2\,\frac{m_\UP m_\down}{m_\UP+m_\down} \,. \nonumber
\ee
The continuity equation and the probability current are still given by (\ref{eq:continu},\ref{eq:current}) provided we define
\be
\XX \ :=\ \left(\sqrt{\frac{m_1}{m}}\ \rr_1, \ \ldots \ , \, \sqrt{\frac{m_N}{m}}\ \rr_N\right)\,.
\label{eq:Ximbal}
\ee

While in the mass-balanced case, there was a single exponent $s$ associated to the unitary three-body problem, in the
mass-imbalanced
case we have two exponents
$s$ and $\tilde{s}$, associated  respectively to the $\UP\UP\down$ and $\UP\down\down$ unitary three-body problems~\cite{Efimov73,Petrov3fermions}.
As a function of $m_\uparrow/m_\downarrow$,
$s$ is continuously decreasing 
and vanishes at the Efimov-effect threshold $m_\uparrow/m_\downarrow = 13.6069 \ldots\ $,
while $\tilde{s}$ is continuously increasing and tends to 2 for $m_\uparrow/m_\downarrow \to \infty$.

\subsection{Extension of the relations from the mass-balanced case} \label{app:sec:gen}

The number of nearby $\UP\UP\down$ triplets is still given by
\be
\boxed{N_{2,1}(\epsilon) 
\underset{\epsilon\to0}{\sim} \,C_{2,1}\ \epsilon^{2s+2}}
\label{eq:N21_imb}
\ee
whereas the number of nearby $\UP\down\down$ triplets is now given by
\be
\boxed{N_{1,2}(\epsilon) 
\underset{\epsilon\to0}{\sim} \,C_{1,2}\ \epsilon^{2\tilde{s}+2}\,.}
\label{eq:N12_imb}
\ee
For our convention $m_\UP > m_\down$, we have $s < \tilde{s}$. Hence the total number of nearby triplets $N_3(\epsilon)$ is dominated by the $\UP\UP\down$ contribution,
so that 
\be
\boxed{N_3(\epsilon) 
\underset{\epsilon\to0}{\sim} \ C_{2,1}\ \epsilon^{2s+2}\,.}
\label{eq:N3_imb}
\ee
The remarks at the end of Sec.~\ref{sec:N3} remain valid.
In particular, 
the bunching effect due
to the zero-range interactions still overcompensates the antibunching effect due to Pauli exclusion,
both for 
$N_{2,1}(\epsilon)$ and $N_{1,2}(\epsilon)$,
because $s < \tilde{s} < 2$.

When three
particles
of spins $\UP, \UP, \down$ approach each other, the asymptotic behavior of the many-body wavefunction is still given by (\ref{eq:R0}).
This yields [using the Jacobian $|\partial(\rr_1,\rr_2,\rr_3) / \partial(\CC,\RR) | = {\rm cos}^3 \varphi\ $]

\be
\boxed{
  C_{2,1} = \sum_{\md=-1}^{+1} C_{2,1}^{(\md)}
}
\nonumber
\ee
with
\be
\boxed{ C_{2,1}^{(\md)} =
N_\UP (N_\UP-1) N_\down
\ \frac{\cos^3\!\varphi}{4\,(s+1)}\ 
\int \left|B_\md(\CC;\rr_4,\ldots,\rr_N)\right|^2
d^3\!C\, \dr_4 \ldots \dr_N.}
\label{eq:C21_BB_imbal}
\ee
When three
particles
of spins $\UP, \down, \down$ approach each other, the asymptotic behavior of the many-body wavefunction is given by (\ref{eq:R0t}) with $s$ replaced by $\tilde{s}$, {\it i.e.}
\be
  \psi(\rr_1,\ldots,\rr_N)
\underset{\tilde{R}\to0}{\sim}
\tilde{R}^{\tilde{s}-2}\!\!\sum_{{\rm m}=-1}^{+1}\!\! \phi_{\rm m}(\tilde{\Oo})\ \tilde{B}_\md(\tilde{\CC};\rr_2,\rr_5,\ldots,\rr_N)\,.
\label{eq:R0t_imbal}
\ee
This yields
\be
\boxed{
  C_{1,2} = \sum_{\md=-1}^{+1} C_{1,2}^{(\md)}
}
\nonumber
\ee
with
\be
\boxed{ C_{1,2}^{(\md)} =
N_\down (N_\down-1) N_\UP
\ \frac{\cos^3\!\tilde{\varphi}}{4\,(\tilde{s}+1)}\ 
\int \left|\tilde{B}_\md(\tilde{\CC}; \rr_2, \rr_5,\ldots,\rr_N)\right|^2
d^3\!\tilde{C}\, \dr_2\,\dr_5 \ldots \dr_N.}
\nonumber
\ee

The triplet correlation functions satisfy
\be
\boxed{\int d^3\!C\ d^5\Omega\ \, g_{2,1}(\rr_1,\rr_2,\rr_3)\underset{R\to0}{\sim}\ 
C_{2,1}\,R^{2s-4}
\ \frac{4(s+1)}{\cos^3\!\varphi}}
\label{eq:g21_C3_imb}
\ee
\be
\boxed{\int d^3\!\tilde{C}\ d^5\tilde{\Omega}\ \, g_{1,2}(\rr_1,\rr_3,\rr_4)\underset{\tilde{R}\to0}{\sim}\ 
C_{1,2}\,\tilde{R}^{2\tilde{s}-4}
\ \frac{4(\tilde{s}+1)}{\cos^3\!\tilde{\varphi}}}
\label{eq:g12_C3_imb}
\ee

The leading large-momentum tail of $\bar{N}_P(K)$ is given by
\be
\boxed{  \bar{N}_P(K) \ \underset{K\to\infty}{\sim} \ \Mr\ \frac{C_{2,1}}{K^{2s+4}} }
\label{eq:NP_tail_imb}
\ee
where $\Mr$ is still given by the expression (\ref{eq:Mr}) in terms of $\Nr$, and $\Nr$ now stands for the normalization constant of the unitary hyperangular wavefunction of the $\UP\UP\down$ problem.\footnote{\bl The contribution from the $\UP\down\down$ three-body problem gives rise to a higher-order subleading tail of $\bar{N}_P(K)$, given by 
$\ds C_{1,2}\ K^{-2\tilde{s}-4}
  \times
32\, \pi^3\,4^{\tilde{s}}\ 3^{-\tilde{s}-1/2}\,(\tilde{s}+1)\,\Gamma(\tilde{s}+2)^2\,\sin^2(\tilde{s}\pi)\,\tilde{\Nr}^2$
with $\tilde{\Nr}$ the normalization constant of the unitary hyperangular wavefunction of the $\UP\down\down$ problem.}

{\gre
The two-particle momentum distribution has the tail
  \be
  \boxed{\lim_{K\to\infty}\ \lim_{k\to\infty} \ K^{2s+4}\ k^4\ \frac{1}{4\pi} \int d\hat{\bf K}\ \ N\left(\frac{m_\UP}{m_\UP+m_\down}\,\KK-{\bf k}\,,\,\frac{m_\down}{m_\UP+m_\down}\,\KK+{\bf k}\right)\ \ = \ \Mr\ C_{2,1}\,.}
  \label{eq:N_k1k3_imb}
  \ee
  }

The expression (\ref{eq:Gamma2,1_real}) for the $\UP\UP\down$ three-body loss rate remains valid. In the expression (\ref{eq:Gamma1,2_real}) for the $\UP\down\down$ three-body loss rate, $s$ is replaced by $\tilde{s}$, {\it i.e.}
$\Gamma_{1,2}  \simeq -\frac{\hbar}{m}\ 8 \tilde{s} (\tilde{s}+1) \, \sum_{\md=-1}^1 C^{(\md)}_{1,2} \ {\rm Im} \,a^{(\md)}_{1,2}$.
Since the $\UP\down\down$ three-body parameters $a^{(\md)}_{1,2}$ are now of order $b^{2\tilde{s}}$ whereas the $\UP\UP\down$ three-body parameters $a^{(\md)}_{2,1}$ are of order $b^{2s}$, the $\UP\down\down$ three-body loss rate  $\Gamma_{1,2} \propto b^{2\tilde{s}}$ is negligible (in the zero-range regime) compared to the $\UP\UP\down$ three-body loss rate $\Gamma_{2,1} \propto b^{2s}$. Hence
\be
\boxed{\Gamma_3 \simeq \Gamma_{2,1} \simeq -\frac{\hbar}{m}\ 8\, s\, (s+1) \, \sum_{\md=-1}^1 C^{(\md)}_{2,1} \ {\rm Im} \,a^{(\md)}_{2,1}.}
\label{eq:Gamma3_imbal}
\ee

The expression (\ref{eq:dE2}) for the two-body contribution to the energy correction remains valid. 
For the 
three-body contribution to the energy correction, we obtain 
\be
\boxed{\delta\!E_3 \ \simeq\ \ 
  \frac{\hbar^2}{m}\ 4\, s\ (s+1)\ \sum_{\md=-1}^1
     C_{2,1}^{(\md)}\ a_{2,1}^{(\md)}
}
\label{eq:dE3_imbal}
\ee
because the leading-order contribution 
again comes from $\UP\UP\down$ triplets.\footnote{\bl 
Indeed, the contribution to $\delta\!E_3$ from $\UP\UP\down$ triplets of nearby particles, given by the r.h.s. of~(\ref{eq:dE3_imbal}), is $\propto b^{2s}$, whereas the $\UP\down\down$ contribution is $\simeq
\frac{\hbar^2}{m}\ \ 4\,  \tilde{s}\ (\tilde{s}+1)\ \ \sum_{\md=-1}^1 C_{1,2}^{(\md)}\ a_{1,2}^{(\md)}\,\propto b^{2\tilde{s}}$.}

A peculiar situation takes place for $s<1/2$ ({\it i.e.} for $m_\UP/m_\down > 12.313\ldots\ $):
$\ \delta\!E_3\propto~\!\!\!b^{2s}$ dominates over $\delta\!E_2 \propto\!\!b\,$;
{\it i.e.}, the finite-range correction mainly comes from configurations with three nearby particles, rather than two nearby particles. This was already pointed out in~\cite{GaoEndoCastin} at the level of the third virial coefficient.
Few-body and many-body numerical computations are often performed with finite-range interactions and extrapolated to the zero-range limit, see {\it e.g.}~\cite{BlumeRelations,Blume3bodyResPRA} and~\cite{zhenyaPRL,zhenyaNJP,zhenyas_crossover,GezerlisXi,CarlsonAFQMC,Carlson_C_relations,ValeC_precise,GandolfiProceeding,KaplanTrap,KaplanXi,Goulko_UFG_2016,SchonenbergConduit,AlhassidPairing,AlhassidC,LeeCondFrac} respectively;
to accurately perform such $b\to0$ extrapolations in the regime $s<1/2$, it may be important to include the $b^{2s}$ scaling.\footnote{\bl Naturally, the $\propto b^{2s}$ scaling may be hard to distinguish from the regular $\propto b$ scaling if $s$ is close to $1/2$ (see {\it e.g.} Fig.~9(b) of \cite{Blume3bodyResPRA}).}$^{,}$\footnote{\bl We take this opportunity to recall two other subtleties relevant to zero-range extrapolations ({\it i.e.} to continuum extrapolations in the case of lattice models, where the interaction range is set by the lattice spacing). Being due to breaking of
Galilean
invariance, they arise for any mass ratio.
The first subtlety, reported in~\cite{zhenyaNJP,LeChapitreIn2,WernerCastinRelationsFermions}, is that for lattice models (where the interaction range $b$ is set by the lattice spacing),
the $\propto r_e$ term given by the r.h.s. of (\ref{eq:dE2})
is not the only $\propto b$ contribution to the two-body finite-range
energy-correction
$\delta\!E_2$:
There is a second contribution to $\delta\!E_2$, proportional to a parameter $R_e$ 
(that parameter $R_e$ quantifies the dependence of the two-body vacuum T-matrix on the center-of-mass momentum, which arises from lattice-induced breaking of Galilean invariance).
Therefore, if one wishes to cancel the $\propto b$ term in the $b\to0$ continuum extrapolation, one needs to tune to zero not only $r_e$ (as done in~\cite{KaplanTrap,KaplanXi,LuuLuscher} and for one of the dispersion relations considered in~\cite{CarlsonAFQMC}) but also $R_e$
(and if the dispersion relation has a cusp at the edge of the Brillouin zone, there is a third $\propto b$ contribution to $\delta\!E_2$,
in a finite box with periodic boundary conditions~\cite{WernerCastinRelationsFermions}).
The second
subtlety,
reported in~\cite{WernerCastinRelationsFermions} and further evidenced in~\cite{AlhassidPairing}, arises if one restricts single-particle momenta to a ball of radius $\Lambda$ (with $\Lambda = \pi/b$ for lattice models, with $b$ the lattice spacing), {\it i.e.} if one takes a dispersion relation equal to $+\infty$ for momenta larger than $\Lambda$, as was done in~\cite{bulgacQMC,BulgacThermodynamics,BulgacCrossover,zhenyas_crossover,BulgacPG,BulgacPG2,BulgacPairing2013,BulgacViscosity}: The universal zero-range model is {\it not} obtained in the limit $\Lambda\to\infty$, because the hard cutoff induces a dependence of the two-body T-matrix on the center-of-mass momentum, and this dependence surprisingly survives for $\Lambda\to\infty$. We expect the same problem in~\cite{KaplanTrap,KaplanXi} where such a spherical cutoff was also used.}

\subsection{Relations within the generalized zero-range model}
\label{app:sec:simpl}  

In this Section, we assume that $m_\UP/m_\down$ is larger than $8.619\ldots\ $,
so that $s < 1$.
This ensures that a three-body wavefunction diverging as $R^{-s-2}$ at small $R$ remains square integrable.
This allows one to define a {\it generalized zero-range model} (GZRM)  by supplementing the two-body contact condition  (\ref{eq:BP})
(involving the two-body scattering length)
by the following three-body contact condition
(involving
the three-body parameters):
\\There exists $B_\md$ such that
\be
\boxed{\psi(\rr_1,\ldots,\rr_N)
 \underset{R\to0}{\sim}  \ 
\ \frac{1}{R^2}
\ \sum_{\md=-1}^{+1} \
\left(
- \frac{a_3^{(\md)}}{R^{s}}
+ R^{s} + o(R^{s})
\right)
\ \phi_\md(\Oo)\ {B}_\md(\CC;\rr_4,\ldots,\rr_N)}
\label{eq:cc_a21}
\ee
in the limit $R\to0$ where particles 1,2,3 approach each other,
while keeping fixed their hyperangles $\Oo$, their center-of-mass $\CC$, and the positions of the other particles $\rr_4, \ldots , \rr_N$.
By antisymmetry,
Eq.~(\ref{eq:cc_a21})
imposes a similar condition when any triplet of particles with spins $\UP\UP\down$ approach
each other.
Apart from the two-body and three-body contact conditions (\ref{eq:BP},\ref{eq:cc_a21}), the GZRM is defined (like the standard zero-range model) by the Schr\"odinger equation without any interaction potential~(\ref{eq:schro_imbal}).
Note that for vanishing three-body parameters, the GZRM reduces to the standard zero-range model (ZRM).

{\bl In the regime where $\ktyp^{\phantom{t}2s}\  |a_3^{(\md)}| \ \gtrsim \ 1$ (for at least one value of $\md$), the ZRM becomes irrelevant, whereas the GZRM remains applicable (provided $\ktyp b \ll 1$). This regime can be reached
\bi
\item
 near a ($\UP\UP\down$)
  three-body resonance \cite{LeChapitreIn2,Petrov3fermions,WernerSym,NishidaSonTan3bRes,Blume3bodyResPRL,GandolfiCarlson3bRes,Blume3bodyResPRA,Sadeghpour_Blume_3bodyRes,Kartavtsev3bRes} 
\item
  when the mass ratio  $m_\UP/m_\down$
  is only slightly smaller than the threshold $13.607\ldots$ where the three-body Efimov effect appears, so that $s$ is small \cite{GaoEndoCastin}.
\ei
}

\vskip 0.12cm
\noindent \underline{\it Remarks:}
\bi
\item
If the underlying microscopic interactions are rotationally-invariant,
 then $a_3^{(\md)}$ does not depend on~$\md$.
\item
  In the present GZRM, in the limit where three particles of spins $\UP\down\down$ approach each other,
the asymptotic behavior of the wavefunction has the same form (\ref{eq:R0t_imbal}) than for the ZRM.
In other words, the behavior $\psi \propto R^{-2-\tilde{s}}$ is forbidden.
Such a wavefunction  would not be a normalizable at small~$R$,
since $\tilde{s}$ is always larger than one.\footnote{\bl Hence there are no $\UP\down\down$ three-body parameters within the GZRM,
  which is why we denoted the $\UP\UP\down$ three-body parameters by $a_3^{(\md)}$ instead of the more cumbersome notation $a_{2,1}^{(\md)}$.}
  This fact does not prevent one from rederiving the relations~(\ref{eq:Gamma3_imbal},\ref{eq:dE3_imbal}), which come from configurations with nearby $\UP\UP\down$ triplets.
  
\item
Extensions of the GZRM to the regime $s\geq 1$, which were introduced recently~\cite{PricoupenkoNbodyRes}, are beyond the scope of this work.

\item
For mathematically rigorous studies of the GZRM in the three-particle case, see~\cite{Minlos3body_res_2014,Teta_3body_res}.

  \ei

\subsubsection{Derivative of the energy with respect to the three-body parameters}
\label{app:imbal,subsec:dE}

Within the GZRM, the derivatives of the energy w.r.t. the three-body
parameters
are given by the three-body contacts,
\be
\boxed{  \frac{\partial E}{\partial a_3^{(\md)}} \ =\ \frac{\hbar^2}{m}\ 4\,s\,( s + 1 )\ C_{2,1}^{(\md)}\, .}
\label{eq:dE_da3}
\ee
Here the derivative w.r.t. 
$a_3^{(\md)}$ is taken at fixed value of $a_2$ and of the other
$a_3^{(\md')}$ with ${\md'\neq\md}$.
\\The three-body contacts $C_{2,1}^{(\md)}$ are still defined by (\ref{eq:C21_BB_imbal}).

\vskip 0.12cm
 \noindent      {\it \underline{Remark:}}\ \ Within the ZRM, the derivative of the energy w.r.t. $a_2$ is given by $C_2$, see~(\ref{eq:dE_da2}).
Within the GZRM, this relation also holds, with $dE/d(-1/a_2)$ replaced by the partial derivative $\partial E/\partial(-1/a_2)$ taken at fixed $a_3^{(\md)}$.
This can be justified by using the derivation presented in Sec.~IV.C of~\cite{WernerCastinRelationsFermions} (at~fixed three-body parameters, there is no additional contribution to the energy variation coming from nearby triplets of particles, as we will see below).

To derive (\ref{eq:dE_da3}) we proceed
as follows.\footnote{\bl This derivation is similar to the case of three bosons treated in App.~A of~\cite{WernerCastinRelationsBosons}.}
We consider two wavefunctions $\psi$ and $\tilde{\psi}$,
which are stationary states of the zero-range model for different
sets of three-body parameters $(a_3^{(\md)})$ and~$(\tilde{a}_3^{(\md)})$,
but the same two-body scattering length $a_2$.
\\Denoting the corresponding eigenenergies by $E$ and $\tilde{E}$, we have
\be \delta := \la \psi, H \tilde{\psi} \ra - \la H \psi, \tilde{\psi} \ra  = (\tilde{E}-E)\,\la \psi| \tilde{\psi} \ra. \nonumber \ee
On the other hand, from the Schr\"odinger equations for $\psi$ and $\tilde{\psi}$,
\be
\delta = - \sum_{i=1}^N \frac{\hbar^2}{2m_i} \int \dr_1 \ldots \dr_N
\, \left(\psi^* \Delta_{\rr_i} \tilde{\psi}
- \tilde{\psi}\, \Delta_{\rr_i} \psi^* \right).
\label{eq:delta_sum_i}
\ee
As we will see below, there is a contribution $\delta_0$ to $\delta$ coming from the configurations where particles 1, 2 and 3 are close to each other.
By symmetry, there is an identical contribution from
the configurations where any set of three particles with spins $\UP\UP\down$
are close to each other.
Hence $\delta$ equals $\delta_0$ times the number of such three-particle sets, $N_\UP (N_\UP-1) N_\down / 2$.
To evaluate $\delta_0$, we only need to keep the terms
$i=1,2,3$ in Eq.~(\ref{eq:delta_sum_i}),
which gives, after
the change of coordinates
$(\rr_1,\rr_2,\rr_3) \longrightarrow (\CC,R,\Oo)$,
\be
\delta_0 = -\frac{\hbar^2}{m}
\cos^3\!\varphi\ 
\int_0^\infty dR\,R^5
\int d^5\Omega
\,\dC
\, \dr_4 \ldots \dr_N
\left\{
\Fr^*
\left(
\frac{\partial^2}{\partial R^2}
+ \frac{1}{R} \frac{\partial}{\partial R}
+ \frac{T_\Oo}{R^2}
+ \frac{\Delta_\CC}{3}
\right)
\tilde{\Fr}
- \left[\Fr^* \leftrightarrow \tilde{\Fr}\right]
\right\}
\label{eq:delta_R}
\ee
where $\Fr := R^2\,\psi$,
$\tilde{\Fr} := R^2\,\tilde{\psi}$,
and $T_\Oo$
is defined by (\ref{eq:def_TOm}).
Since the two-body scattering length is the same for $\psi$ and $\tilde{\psi}$,
we only need to keep the terms involving derivatives with respect to $R$ in Eq.~(\ref{eq:delta_R}).\footnote{\bl For a more detailed justification of this step, see the reasoning around Eq.~(A7) of~\cite{WernerCastinRelationsBosons}.}
Transforming the integral over $R$ into a boundary term at $R\to0$, we then get
\be
\delta_0 = \frac{\hbar^2}{m}
\cos^3\!\varphi\,
\int \,\dC
\, \dr_4 \ldots \dr_N
\,\int d^5\Omega
\ \lim_{R\to0}
R \left(
\Fr^* \ \frac{\partial \tilde{\Fr}}{\partial R}
-
\tilde{\Fr} \ \frac{\partial \Fr^*}{\partial R}
\right).
\nonumber
\ee
The result (\ref{eq:dE_da3}) follows by using the three-body contact conditions [given by Eq.~(\ref{eq:cc_a21}) for $\psi$, and the same condition with $\tilde{a}_3^{(\md)}$ for $\tilde{\psi}$], taking the limit $\tilde{a}_3^{(\md)}  \to {a}_3^{(\md)}$, and using the expression~(\ref{eq:C21_BB_imbal}) of the three-body contacts.

\subsubsection{Three-body loss rate}

Let us consider the GZRM with complex three-body parameters, $a_3^{(\md)}\in\mathbb{C}$. The three-body loss rate is then given by
\be
\boxed{\Gamma_3 = -\frac{\hbar}{m}\ 8\, s\, (s+1) \, \sum_{\md=-1}^1 C^{(\md)}_{2,1} \ {\rm Im} \,a^{(\md)}_{3}.}
\label{eq:Gamma3_GZRM}
\ee
Within the GZRM, there are only $\UP\UP\down$ losses, and no $\UP\down\down$ losses, {\it i.e.} $\Gamma_3 = \Gamma_{2,1}$ and $\Gamma_{1,2}=0$.
\\To derive this relation, we proceed very similarly to the bosonic case of~\cite{WernerCastinRelationsBosons}.
We consider a stationary state of the GZRM, {\it i.e.} a solution $\psi(\rr_1,\ldots,\rr_N)$ of the stationary Schr\"odinger equation~(\ref{eq:schro_imbal}) together with the two-body and three-body contact conditions (\ref{eq:BP},\ref{eq:cc_a21}).
The corresponding solution of the time-dependent Schr\"odinger equation is $\Psi(\rr_1,\ldots,\rr_N;t) = \psi(\rr_1,\ldots,\rr_N) e^{-i E t / \hbar}$, and $\Gamma_3 = - \left.\frac{\partial}{\partial t}\right|_{t=0} \int \dr_1 \ldots \dr_N \, \left|\Psi(\rr_1,\ldots,\rr_N;t)\right|^2$, with $\psi$ normalized to unity.
Excluding from the
integration
domain the regions where $R_{ijk}<\epsilon$ and taking the limit $\epsilon\to0$, the continuity equation and the three-body contact conditions leads to the final expression,
where all extra mass-ratio dependent factors, arising {\it e.g.} from (\ref{eq:Ximbal}), divide out. 

}

{\gre
\subsection{Validity conditions} 
\label{sec:max_ratio}

Let us denote by $s_{j_\UP,j_\down}$ the scaling exponent of the unitary $(j_\UP+j_\down)$ body problem with $j_\UP$ particles of spin $\UP$ and $j_\down$ particles of spin $\down$
(so that $s_{2,1} \equiv s$ and $s_{1,2} \equiv \tilde{s}$\,).
We expect the relations (\ref{eq:N21_imb},\ref{eq:N3_imb},\ref{eq:NP_tail_imb},\ref{eq:N_k1k3_imb}) to be valid under the condition ${\rm Min} \left\{ s_{j_\UP,j_\down}; \ j_\UP \geq 2, \ j_\down \geq 1, \ j_\UP + j_\down > 3 \right\} > s$.
Indeed, we expect a contribution from the $(j_\UP+j_\down)$ body problem to $N_{2,1}(\epsilon)$ at small $\epsilon$ [resp. to $\bar{N}_P(K)$ at large $K$] scaling as $\epsilon^{2s_{j_\UP,j_\down}+2}$ \big(resp.~$1/K^{2s_{j_\UP,j_\down}+4}$\,\big),
which dominates over the contribution from the three-body problem when $s_{j_\UP,j_\down}<s$.
Similarly, we expect relation~(\ref{eq:N12_imb}) to be valid under the condition ${\rm Min} \left\{ s_{j_\UP,j_\down}; \ j_\UP \geq 1, \ j_\down \geq 2, \ j_\UP + j_\down > 3 \right\}>\tilde{s}$.
The former condition breaks down when $m_\UP/m_\down$ is near the thresholds where the four-body~\cite{CMP} and five-body~\cite{PetrovBazak5body} Efimov effects appear, where $s_{3,1}$ and $s_{4,1}$ become small.
Based on existing data, we expect that 
both of the above validity conditions are satisfied
at least in the range $m_\UP/m_\down \leq 10$.
Indeed,
in this range,
we have $s_{3,1}>s$ and $s_{4,1}>s$~\cite{PetrovBazak5body},
$s_{5,1}>s$~\cite{Bazak6body},
$s_{2,2}>\tilde{s}$~\cite{Blume3bodyResPRA},
$s_{1,3} > \tilde{s}$~\cite{PetrovBazak5body},
and the trends of available data suggest that the conditions will also hold for larger values of $j_\UP$ or $j_\down$.
We conservatively restricted to $m_\UP/m_\down \leq 10$ because $s_{2,2}$ was not computed beyond this range,
but the conditions $s_{3,1}>s$ and $s_{4,1}>s$ actually hold up to at least $m_\UP/m_\down = 13.2$~\cite{PetrovBazak5body}.


}

\section{Generalization to statistical mixtures and non-stationary states}
\label{app:non_stat}

{\bl Many of} the relations derived for stationary states in the main text {\bl and in Appendix~\ref{app:imbal}} are directly generalizable to non-stationary states and statistical mixtures,
similarly to the relations involving the two-body contact~\cite{TanEnergetics,TanLargeMomentum,WernerCastinRelationsFermions}.
Indeed, Eqs.~(\ref{eq:R0},\ref{eq:C3_BB},\ref{eq:R0t},\ref{eq:C3_BBt}) remain valid for any
non-pathological linear combinations of stationary states
(including solutions of time-dependent problems, where $a_2$ and the trapping potential $U$ can depend on time);
and relations~(\ref{eq:N3},\ref{eq:N21},\ref{eq:N12},\ref{eq:g21_C3},\ref{eq:g12_C3},\ref{eq:NP_tail}) remain true for arbitrary non-pathological statistical mixtures of such pure states
(including the case of thermal equilibrium).
Here, non-pathological means that the
occupation probabilities of  stationary states should decay sufficiently quickly (which includes the simple case where only a finite number of states are populated);
more specifically,
for a pure state,
it is necessary
that  Eqs.~(\ref{eq:BP},\ref{eq:R0_b}) still hold,
while
for a statistical mixture
$\hat{\rho} = \sum_i c_i |\psi_i\rangle \langle \psi_i |$,
it is necessary that
the three-body contact of the mixture
[defined by Eq.~(\ref{eq:N3})]
equals
$\sum_i c_i \,C_{3,i}$ with $C_{3,i}$ the three-body contact of state~$\psi_i$.

{\bl Moreover, at thermal equilibrium, 
  the thermally averaged loss rates are given by~(\ref{eq:Gamma3},\ref{eq:Gamma2,1},\ref{eq:Gamma1,2}) for simple interactions; for more general interactions they are given by~(\ref{eq:Gamma2,1_real},\ref{eq:Gamma1,2_real}), and by (\ref{eq:Gamma3_real}) for a homogeneous unpolarized gas.
  Furthermore, (\ref{eq:dE_21_12},\ref{eq:dE3}) remain valid in the canonical ensemble, with $\delta\!E_3$ the energy difference (between finite-range and zero-range models) taken at fixed entropy (which equals the free-energy difference at fixed temperature).
}

As for the wavevector $\ktyp$ that appears in the validity conditions,
it should 
be defined by the same procedure as before and then taking the maximum over all
significantly populated eigenstates; for example for the balanced {\bl unitary} gas at thermal equilibrium, $\ktyp = {\rm max}(k_F, k_T)$
with $k_T$ the thermal wavevector, defined by $\hbar^2 k_T^2 / (2 m) = k_B T$.

\bibliographystyle{crunsrt}


\bibliography{felix_copy}

\end{document}